
 %
\input harvmac
 %
\catcode`@=11
\def\rlx{\relax\leavevmode}                  
 %
 %
 %
\font\tenmib=cmmib10
\font\sevenmib=cmmib10 at 7pt 
\font\fivemib=cmmib10 at 5pt  
\font\tenbsy=cmbsy10
\font\sevenbsy=cmbsy10 at 7pt 
\font\fivebsy=cmbsy10 at 5pt  
\def\BMfont{\textfont0\tenbf \scriptfont0\sevenbf
                              \scriptscriptfont0\fivebf
            \textfont1\tenmib \scriptfont1\sevenmib
                               \scriptscriptfont1\fivemib
            \textfont2\tenbsy \scriptfont2\sevenbsy
                               \scriptscriptfont2\fivebsy}
\def\BM#1{\rlx\ifmmode\mathchoice
                      {\hbox{$\BMfont#1$}}
                      {\hbox{$\BMfont#1$}}
                      {\hbox{$\scriptstyle\BMfont#1$}}
                      {\hbox{$\scriptscriptstyle\BMfont#1$}}
                 \else{$\BMfont#1$}\fi}
 %
 %
 %
 %
\def\inbar{\vrule height1.5ex width.4pt depth0pt}
\def\sinbar{\vrule height1ex width.35pt depth0pt}
\def\ssinbar{\vrule height.7ex width.3pt depth0pt}
\font\cmss=cmss10
\font\cmsss=cmss10 at 7pt
\def\ZZ{\rlx\leavevmode
             \ifmmode\mathchoice
                    {\hbox{\cmss Z\kern-.4em Z}}
                    {\hbox{\cmss Z\kern-.4em Z}}
                    {\lower.9pt\hbox{\cmsss Z\kern-.36em Z}}
                    {\lower1.2pt\hbox{\cmsss Z\kern-.36em Z}}
               \else{\cmss Z\kern-.4em Z}\fi}
\def\Ik{\rlx{\rm I\kern-.18em k}}  
\def\IC{\rlx\leavevmode
             \ifmmode\mathchoice
                    {\hbox{\kern.33em\inbar\kern-.3em{\rm C}}}
                    {\hbox{\kern.33em\inbar\kern-.3em{\rm C}}}
                    {\hbox{\kern.28em\sinbar\kern-.25em{\sevenrm C}}}
                    {\hbox{\kern.25em\ssinbar\kern-.22em{\fiverm C}}}
             \else{\hbox{\kern.3em\inbar\kern-.3em{\rm C}}}\fi}
\def\IP{\rlx{\rm I\kern-.18em P}}
\def\IR{\rlx{\rm I\kern-.18em R}}
\def\Ione{\rlx{\rm 1\kern-2.7pt l}}
 %
 %

 %

\def\intem#1{\par\leavevmode%
              \llap{\hbox to\parindent{\hss{#1}\hfill~}}\ignorespaces}
 %


 %
\newskip\humongous \humongous=0pt plus 1000pt minus 1000pt   
\def\caja{\mathsurround=0pt}
\newif\ifdtup
 %
\def\eqalign#1{\,\vcenter{\openup2\jot \caja
     \ialign{\strut \hfil$\displaystyle{##}$&$
      \displaystyle{{}##}$\hfil\crcr#1\crcr}}\,}
 %

 %
\def\panorama{\global\dtuptrue \openup2\jot \caja
     \everycr{\noalign{\ifdtup \global\dtupfalse
      \vskip-\lineskiplimit \vskip\normallineskiplimit
      \else \penalty\interdisplaylinepenalty \fi}}}
 %
\def\eqalignno#1{\panorama \tabskip=\humongous
     \halign to\displaywidth{\hfil$\displaystyle{##}$
      \tabskip=0pt&$\displaystyle{{}##}$\hfil
       \tabskip=\humongous&\llap{$##$}\tabskip=0pt\crcr#1\crcr}}
 %
\def\eqalignnotwo#1{\panorama \tabskip=\humongous
     \halign to\displaywidth{\hfil$\displaystyle{##}$
      \tabskip=0pt&$\displaystyle{{}##}$
       \tabskip=0pt&$\displaystyle{{}##}$\hfil
        \tabskip=\humongous&\llap{$##$}\tabskip=0pt\crcr#1\crcr}}
 %

 %

 %
 %
 %
 %
\let\lv=\l          
\let\ii=\i          
\def\,{\hskip1.5pt}           
 %
\let\a=\alpha
\let\b=\beta
\let\c=\chi
\let\d=\delta       \let\vd=\partial             \let\D=\Delta
\let\e=\epsilon     
\let\f=\phi         \let\vf=\varphi              \let\F=\Phi
\let\g=\gamma                                    
\let\h=\eta
\let\i=\iota
\let\j=\psi                                      
\let\k=\kappa
\let\l=\lambda                                   
\let\m=\mu
\let\n=\nu
\let\p=\pi                         
\let\q=\theta       \let\vq=\vartheta            \let\Q=\Theta
\let\r=\rho         
\let\s=\sigma                   \let\S=\Sigma

\let\w=\omega                                    \let\W=\Omega

 %
 %
\def\Box{\sqcap\llap{$\sqcup$}}
\def\lapp{\lower.4ex\hbox{\rlap{$\sim$}} \raise.4ex\hbox{$<$}}
\def\gapp{\lower.4ex\hbox{\rlap{$\sim$}} \raise.4ex\hbox{$>$}}
\def\con{\ifmmode\raise.1ex\hbox{\bf*}
          \else\raise.1ex\hbox{\bf*}\fi}
\def\bo{{\raise.15ex\hbox{\large$\Box\kern-.39em$}}}

\let\into=\hookrightarrow

\def\dual{\relax\leavevmode\lower.9ex\hbox{\titlerms*}}
\def\define{\buildrel\rm def\over =}
\let\id=\equiv
\let\8=\otimes
 %
 %
 %
 %
\let\ba=\overline
\let\2=\underline
\let\ha=\widehat
\let\Tw=\widetilde
 %
\def\dt#1{{\buildrel{\smash{\lower1pt\hbox{.}}}\over{#1}}}

\font\eightrm=cmr8
\def\6(#1){\relax\leavevmode\hbox{\eightrm(}#1\hbox{\eightrm)}}
\def\0#1{\relax\ifmmode\mathaccent"7017{#1}     
                \else\accent23#1\relax\fi}      
\def\7#1#2{{\mathop{\null#2}\limits^{#1}}}      
\def\5#1#2{{\mathop{\null#2}\limits_{#1}}}      
 %
\def\bra#1{\left\langle #1\right|}
\def\ket#1{\left| #1\right\rangle}
\def\V#1{\langle#1\rangle}

 %

 %

 %

 %
\newbox\t@b@x
\def\rightarrowfill{$\m@th \mathord- \mkern-6mu
     \cleaders\hbox{$\mkern-2mu \mathord- \mkern-2mu$}\hfill
      \mkern-6mu \mathord\rightarrow$}
\def\tooo#1{\setbox\t@b@x=\hbox{$\scriptstyle#1$}%
             \mathrel{\mathop{\hbox to\wd\t@b@x{\rightarrowfill}}%
              \limits^{#1}}\,}
\def\leftarrowfill{$\m@th \mathord\leftarrow \mkern-6mu
     \cleaders\hbox{$\mkern-2mu \mathord- \mkern-2mu$}\hfill
      \mkern-6mu \mathord-$}
\def\froo#1{\setbox\t@b@x=\hbox{$\scriptstyle#1$}%
             \mathrel{\mathop{\hbox to\wd\t@b@x{\leftarrowfill}}%
              \limits^{#1}}\,}
 %
\def\frac#1#2{{#1\over#2}}
\def\frc#1#2{\relax\ifmmode{\textstyle{#1\over#2}} 
                    \else$#1\over#2$\fi}           
\def\inv#1{\frc{1}{#1}}                            
 %
\def\Claim#1#2#3{\bigskip\begingroup%
                  \xdef #1{\secsym\the\meqno}%
                   \writedef{#1\leftbracket#1}%
                    \global\advance\meqno by1\wrlabeL#1%
                     \noindent{\bf#2}\,#1{}\,:~\sl#3\vskip1mm\endgroup}

\def\QED{\rlx\hfill$\Box$\kern-7pt\raise3pt\hbox{$\surd$}\bigskip}
 %
 %

\def\K#1#2{\relax\def\normalbaselines{\baselineskip12pt\lineskip3pt
                                       \lineskiplimit3pt}
             \left[\matrix{#1}\right.\!\left\|\,\matrix{#2}\right]}
\def\muthstrut{\vphantom1}
\def\mutrix#1{\null\,\vcenter{\normalbaselines\m@th
        \ialign{\hfil$##$\hfil&&~\hfil$##$\hfill\crcr
            \muthstrut\crcr\noalign{\kern-\baselineskip}
            #1\crcr\muthstrut\crcr\noalign{\kern-\baselineskip}}}\,}

 %
\def\YT#1#2{\vcenter{\hbox{\vbox{\baselineskip=\normalbaselineskip%
             \def\Box{$\sqcap\llap{$\sqcup$}$\kern-1.2pt}%
              \def\Z{\hfil\vskip-12pt}%
               \setbox0=\hbox{#1}\hsize\wd0\parindent=0pt#2}\,}}}
\def\EU{\rlx\ifmmode \c_{{}_E} \else$\c_{{}_E}$\fi}
\def\TM{\rlx\ifmmode {\cal T_M} \else$\cal T_M$\fi}
\def\TW{\rlx\ifmmode {\cal T_W} \else$\cal T_W$\fi}
\def\CM{\rlx\ifmmode {\cal T\rlap{\bf*}\!\!_M}
             \else$\cal T\rlap{\bf*}\!\!_M$\fi}
\def\hm#1#2{\rlx\ifmmode H^{#1}({\cal M},{#2})
                 \else$H^{#1}({\cal M},{#2})$\fi}
\def\CP#1{\rlx\ifmmode\IP^{#1}\else\IP$^{#1}$\fi}
\def\cP#1{\rlx\ifmmode\IC{\rm P}^{#1}\else$\IC{\rm P}^{#1}$\fi}

\def\sll#1{\rlx\rlap{\,\raise1pt\hbox{/}}{#1}}
\def\Sll#1{\rlx\rlap{\,\kern.6pt\raise1pt\hbox{/}}{#1}\kern-.6pt}
%

 %
 %
\def\CFT{conformal field theory}
\def\CFTs{conformal field theories}
\def\CY{Calabi-\kern-.2em Yau}
\def\LGO{Landau-Ginzburg orbifold}
\def\3{\ifmmode\ldots\else$\ldots$\fi}
\def\Z{\hfil\break\rlx\hbox{}\quad}
\def\3{\ifmmode\ldots\else$\ldots$\fi}
\def\?{d\kern-.3em\raise.64ex\hbox{-}}           
\def\9{\raise.43ex\hbox{-}\kern-.37em D}         
\def\ping{\nobreak\par\centerline{---$\circ$---}\goodbreak\bigskip}
 %
 %
\def\I#1{{\it ibid.\,}{\bf#1\,}}
\def\Pre#1{{\it #1\ University report}}
\def\pre#1{{\it University of #1 report}}

\def\NP#1{{\it Nucl.\,Phys.\,}{\bf#1\,}}
\def\PL#1{{\it Phys.\,Lett.\,}{\bf#1\,}}

\def\MPL#1{{\it Mod.\,Phys.\,Lett.\,}{\bf#1\,}}

\def\CMP#1{{\it Commun.\,Math.\,Phys.\,}{\bf#1\,}}
\def\CQG#1{{\it Class.\,Quant.\,Grav.\,}{\bf#1\,}}

 %
 %
 %
\baselineskip=13.0861pt plus2pt minus1pt
\parskip=\medskipamount
\let\ft=\foot
\noblackbox
\def\SaveTimber{\abovedisplayskip=1.5ex plus.3ex minus.5ex
                \belowdisplayskip=1.5ex plus.3ex minus.5ex
                \abovedisplayshortskip=.2ex plus.2ex minus.4ex
                \belowdisplayshortskip=1.5ex plus.2ex minus.4ex
                \baselineskip=12pt plus1pt minus.5pt
 \parskip=\smallskipamount
 \def\ft##1{\unskip\,\begingroup\footskip9pt plus1pt minus1pt\setbox%
             \strutbox=\hbox{\vrule height6pt depth4.5pt width0pt}%
              \global\advance\ftno by1\footnote{$^{\the\ftno)}$}{##1}%
               \endgroup}
 \def\listrefs{\footatend\vfill\immediate\closeout\rfile%
                \writestoppt\baselineskip=10pt%
                 \centerline{{\bf References}}%
                  \bigskip{\frenchspacing\parindent=20pt\escapechar=` %
                   \rightskip=0pt plus4em\spaceskip=.3333em%
                    \input \jobname.refs\vfill\eject}\nonfrenchspacing}}
 %
\def\Afour{\ifx\answ\bigans
            \hsize=16.5truecm\vsize=24.7truecm
             \else
              \hsize=24.7truecm\vsize=16.5truecm
               \fi}
\catcode`@=12
 %
 %
\def\rd{{\rm d}}
\def\IF{\rlx{\rm I\kern-.18em F}}
\def\pmod#1{\allowbreak\mkern8mu({\rm mod}\,#1\,)}
\def\db{\relax\leavevmode\hbox{$\partial$\kern-.4em\vrule
         height1.86ex depth-1.8ex width4pt}\kern.8pt\vphantom{\vd}}
\def\lb{\relax\leavevmode\hbox{$\l$\kern-.4em\vrule
         height1.86ex depth-1.8ex width4pt}\kern.8pt\vphantom{\l}}
\def\qb{\relax\leavevmode\hbox{$q$\kern-.4em\vrule
         height1.26ex depth-1.2ex width4pt}\kern.8pt\vphantom{q}}
\def\Cb{\relax\leavevmode\hbox{$C$\kern-.53em\vrule
         height1.9ex depth-1.8ex width5pt}\kern.5pt\vphantom{C}}
\def\Db{\relax\leavevmode\hbox{$D$\kern-.6em\vrule
         height1.9ex depth-1.8ex width5pt}\kern.5pt\vphantom{D}}
\def\Gb{\relax\leavevmode\hbox{$G$\kern-.53em\vrule
         height1.9ex depth-1.8ex width5pt}\kern.5pt\vphantom{G}}
\def\Jb{\relax\leavevmode\hbox{$J$\kern-.4em\vrule
         height1.9ex depth-1.8ex width4.5pt}\kern.5pt\vphantom{J}}
\def\Kb{\relax\leavevmode\hbox{$K$\kern-.65em\vrule
         height1.9ex depth-1.8ex width5.5pt}\kern.5pt\vphantom{K}}
\def\Lb{\relax\leavevmode\hbox{$L$\kern-.5em\vrule
         height1.9ex depth-1.8ex width5pt}\kern.0pt\vphantom{L}}
\def\Nb{\relax\leavevmode\hbox{$N$\kern-.6em\vrule
         height1.9ex depth-1.8ex width5pt}\kern.0pt\vphantom{N}}
\def\Pb{\relax\leavevmode\hbox{$P$\kern-.55em\vrule
         height1.9ex depth-1.8ex width4.5pt}\kern.5pt\vphantom{P}}
\def\Qb{\relax\leavevmode\hbox{$Q$\kern-.55em\vrule
         height1.9ex depth-1.8ex width5.5pt}\kern.5pt\vphantom{Q}}
\def\Tb{\relax\leavevmode\hbox{$T$\kern-.6em\vrule
         height1.9ex depth-1.8ex width6pt}\kern.0pt\vphantom{T}}
\def\Xb{\relax\leavevmode\hbox{$X$\kern-.6em\vrule
         height1.9ex depth-1.8ex width5pt}\kern.5pt\vphantom{X}}
\def\XB{\relax\leavevmode\hbox{$\BM{X}$\kern-.65em\vrule
         height1.95ex depth-1.85ex width6pt}\kern.5pt\vphantom{X}}
\def\Ssl{\rlap{\kern1.2pt\raise1pt\hbox{\rm/}}{\hbox{$S$}}}
\def\Afour{\ifx\answ\bigans\hsize=16.5truecm\vsize=24.7truecm
                      \else\hsize=24.7truecm\vsize=16.5truecm\fi}
\SaveTimber     
 %
 %
 %
 %
\Title{\vbox{\baselineskip12pt \hbox{HUPAPP-93/2}
                                  \hbox{NSF-ITP-93-37}
                                  \hbox{UTTG-10-93}}}
      {\vbox{\centerline{Couplings for Compactification}}}
\centerline{\titlerms Per Berglund\footnote{$^{\diamondsuit}$}
            {After Sept.~15th: Institute for Advanced Study,
             Olden Lane, Princeton, NJ 08540, USA.}}      \vskip 0mm
 \centerline{\it Institute for Theoretical Physics}       \vskip-1mm
 \centerline{\it University of California}
                                                          \vskip-1mm
 \centerline{\it Santa Barbara, CA 93106}                 \vskip 0mm
 \centerline{and}                                         \vskip 0mm
 \centerline{\it Theory Group, Department of Physics}     \vskip-1mm
 \centerline{\it University of Texas, Austin, TX 78712}   \vskip-1mm
 \centerline{\rm berglund\,@\,sbitp.ucsb.edu}              \vskip 0mm
\vskip .2in
 \centerline{\titlerms and}
\vskip .2in
\centerline{\titlerms Tristan H\"ubsch\footnote{$^{\spadesuit}$}
            {On leave from the Institute ``Ru\?er
             Bo\v skovi\'c'', Zagreb, Croatia.}}          \vskip0mm
 \centerline{\it Department of Physics}                   \vskip-1mm
 \centerline{\it Howard University, Washington, DC~20059} \vskip-1mm
 \centerline{\rm hubsch\,@\,reliant.cldc.howard.edu}
\vfill

\centerline{ABSTRACT}\vskip2mm
\vbox{\narrower\narrower\narrower\baselineskip=12pt\noindent
A general formula is obtained for Yukawa couplings in compactification
on \LGO{s} and corresponding \CY\ spaces. Up to the kinetic term
normalizations, this equates the classical Koszul ring structure with
the \LGO\ chiral ring structure and the true super\CFT\ ring
structure.}

\Date{\vbox{ \line{3/\number\yearltd \hfill}}}
\footline{\hss\tenrm--\,\folio\,--\hss}
 %
 %
 %
\nref\rPCSW{P.~Candelas and S.~Weinberg: \NP{B237}(1984)397.}

\nref\rPhilip{P.~Candelas: \NP{B298}(1988)458. }

\nref\rYukEx{J.~Distler and B.R.~Greene: \NP{B309}(1988)295\semi
       B.R.~Greene, C.A.~L\"utken and G.G.~Ross:
       \NP{B325}(1989)101\semi
       S.F.~Cordes and Y.~Kikuchi: Correlation Functions and
       Selection Rules in Minimal N=2 String Compactifications.
       \pre{Texas A\&M} CTP-TAMU-92/88 (1988, unpublished),
       \MPL{A4}(1989)1365\semi
       R.Schimmrigk: \PL{229B}(1989)227.}

\nref\rCYCI{T.~H\"ubsch: \CMP{108}(1987)291\semi
       P.~Green and T.~H\"ubsch: \CMP{109}(1987)99\semi
       P.~Candelas, A.M.~Dale, C.A.~L\"utken and R.~Schimmrigk:
       \NP{B298}(1988)493.}

\nref\rBeast{T.~H\"ubsch: {\it \CY\ Manifolds---A Bestiary for
       Physicists}\Z (World Scientific, Singapore, 1992).}

\nref\rWCP{P.~Candelas, M.~Lynker and R.~Schimmrigk:
       \NP{B341}(1990)383\semi
       A.~Klemm and R.~Schimmrigk~: Landau-Ginzburg String Vacua,
       CERN preprint CERN-TH 6459/92\semi
       M.~Kreuzer and H.~Skarke: No Mirror Symmetry in
       Landau-Ginzburg Spectra, CERN preprint CERN-TH 6461/92.}

\nref\rMichael{M.G.~Eastwood: {\it
       Math.\,Proc.\,Camb.\,Phil.\,Soc.\,}{\bf97}(1985)165\semi
       M.G.~Eastwood and T.~H\"ubsch: \CMP{132}(1990)383.}

\nref\rMatter{P.~Berglund, T.~H\"ubsch and L.~Parkes:
       \MPL{A5}(1990)1485, \CMP{148}(1992)57.}

\nref\rBVW{B.R.~Greene, C.~Vafa and N.P.~Warner:
       \NP{B324}(1989)371.}

\nref\rWBRG{B.R.~Greene: \CMP{130}(1990)335.}

\nref\rElusive{T.~H\"ubsch: \CQG{8}(1991)L31.}

\nref\rCQG{P.~Berglund, B.~Greene and T.~H\"ubsch:
       \MPL{A7}(1992)1855.}

\nref\rLGO{C.~Vafa: \MPL{A4}(1989)1169\semi
       K.~Intrilligator and C.~Vafa: \NP{B339}(1990)95.}

\nref\rChiRi{C.~Vafa and N.~Warner: \PL{218B}(1989)51\semi
       W.~Lerche, C.~Vafa and N.~Warner: \NP{B324}(1989)427.}

\nref\rSL{T.~H\"ubsch and S.-T.~Yau: \MPL{A7}(1992)3277;
       see also {\it Essays on Mirror Manifolds}~p.372--387, ed.\
       S.-T.~Yau, (International Press, Hong Kong, 1992).}

\nref\rGepnerIII{D.~Gepner: String Theory on \CY\ Manifolds:
       the Three Generations Case. \Pre{Princeton} (December 1987,
       unpublished).}

\nref\rHartshorne{R.~Hartshorne: {\it Algebraic Geometry}
       (Springer-Verlag, New York, 1977).}

\nref\rGepner{D.~Gepner: \PL{199B}(1987)380.}

\nref\rRolf{R.~Schimmrigk: \PL{193B}(1987)175.}

\nref\rCeGiPa{S.~Cecotti, L.~Girardello and A.~Pasquinucci:
       \NP{B328}(1989)701.}

\nref\rPDM{P.~Green and T.~H\"ubsch: \CMP{113}(1987)505.}

\nref\rDGNR{J.~Distler and B.R.~Greene:
       \NP{B309}(1988)295.}

\nref\rGCM{P.~Berglund and T.~H\"ubsch: \NP{B}(in press), also in
      {\it Essays on Mirror Manifolds}~p.388--407, ed.\ S.-T.~Yau,
       (International Press, Hong Kong, 1992).}

\nref\rCubi{J.~Distler, B.~Greene, K.~Kirklin and P.~Miron:
       \PL{195B}(1987)41\semi
       P.~Green and T.~H\"ubsch: \CQG{6}(1989)311.}

\nref\rDSWW{M.~Dine, N.~Seiberg, X.G.~Wen and E.~Witten:
       \NP{B278}(1986)769, \I{B289}(1987)319.}

\nref\rMax{M.~Kreuzer and H.~Skarke: On the Classification of
       Quasihomogeneous Functions, CERN preprint CERN-TH-6373/92.}

%
 %
\newsec{Introduction, Results and Synopsis}\noindent
In {\it any} {Ka\lv}u\.za-Klein type compactification on a compact
`internal space' $\cal M$, the coupling parameters of the effective
spacetime field theory depend on the geometry of this `internal
space' and can often be determined from it~\refs{\rPCSW, \rPhilip}.
Even when a geometrical description of this `internal sector' eludes
us, valuable selection rules and sometimes even exact and complete
solutions can be obtained; see for example, Refs.~\rYukEx.

Consider superstring \CY\ compactifications, that is, superstring
models with an `internal sector' $\cal M$ which is known to
correspond to a compact, complex 3-fold with trivial canonical class.
Superstring propagation is governed by a $\s$-model which may be
viewed as (a)~describing the world sheet $\S$ immersed in $\cal M$ or
(b)~a 2-dimensional quantum field theory on $\S$. The geometry of maps
$\S \into \cal M$ is then promisingly linked to the quantum field
theory aspects of such models as one can use the methods in one field
to obtain results in the other; correlation functions of the
2-dimensional field theory are couplings of the compactified effective
spacetime field theory.

Clearly, the analysis presented here may also be regarded as a toy
model: the geometry of the target $\cal M$ and the configuration space
$\{\S \into{\cal M}\}$ of {\it any} $\s$-model may be employed to
advance our understanding of quantum field theory. In this sense, our
results pertain to the algebra of observables in a wide class of
quantum field theories. Since the algebraic geometry which we use
remains valid regardless of the triviality of the canonical class
(criticality), the relations and applications to field theory can be
made equally general, although perhaps not as easily interpreted. For
sake of immediate application, we do focus on (2,2)-supersymmetric,
that is, \CY\ compactifications only and will confine our
cross-dictionary accordingly. \ping

Take, for instance, $\cal M$ to be a \CY\ complete intersection of
hypersurfaces in a product of flag-spaces $\IF_{\vec{n}} \define
U(N)/\big( \prod_i U(n_i)\big)$, where $N=\sum_i n_i$. Already the
simplest case involving products of projective spaces $\IP^n =
U(n{+}1)/ \big(U(1) \times U(n)\big)$ provides several thousand
examples~\rCYCI.
The weighted (quasi-homogeneous) variants are obtained as quotients
$\IF_{\vec{n}}/{\mit\D}_{\vec{m}}$, with quasi-homogeneous coordinates
$y_i \define x_i{}^{m_i}$, and where
$\mit\D_{\vec{m}}= \prod_i \ZZ_{m_i}$
is the cyclic group which leaves the $y_i$ invariant~\rBeast. Again,
even just simple hypersurfaces in a single weighted projective space
provide several thousand examples~\rWCP.

For all such $\cal M$ (in fact, {\it regardless} of being \CY), all
desired cohomology may be computed using equivariant (co)homological
algebra based on the Koszul sequence and the Bott-Borel-Weil
Theorem~\refs{\rBeast, \rMichael, \rMatter}; call this simply `Koszul
computation'.

A favorable subset of such manifolds define \LGO{s} and thereby
super\CFTs~\refs{\rBVW, \rWBRG} as the limit of their renormalization
flow; see however also Refs.~\refs{\rElusive, \rCQG}. The {\bf27} and
{\bf27*} massless fields in such models may be found by semiclassical
methods~\rLGO\ and correspond to charge-$(1,1)$ and -$({-}1,1)$
states in the $(2,2)$-supersymmetric field theory.  Note that each
flag-space may be identified with a quotient space other than in the
definition above: for example, $\IP^n = U(n{+}1)/ \big(U(1) \times
U(n)\big)$ also equals the quotient $\IP^n = \IC^{n+1}/\IC^*$, where
$\IC^*$ is the multiplicative group by non-zero complex numbers. The
defining polynomial of the hypersurface $f(x)=0$ in $\IP^n$ becomes
the superpotential of the Landau-Ginzburg field theory, with the
fields spanning the {\it affine} space $\IC^{n+1}$; the quotient
thereof is the \LGO\ft{To find the spectrum, passing to a quotient by
a finite subgroup of $\IC^*$ suffices.} which corresponds to the
projective space ${\cal M} \into \IP^n$. Thus, we write
$\IP^4_{(w_1,\ldots,w_5)}[d]$ for the family of degree-$d$
hypersurfaces in the weighted projective 4-space with weights $w_i$,
but $\IC^5_{(w_1,\ldots,w_5)}[d]$ for the affine configuration space
of the corresponding Landau-Ginzburg field theory.

It was proved~\rCQG\ that {\it beyond the final agreement} of the
spectra as found, respectively, by the Koszul and the \LGO\
calculations, the two approaches display a remarkably precise
correspondence in many details. This is most easily stated as being a
(somewhat formal) isomorphism between the `Chiral Ring' structure of
\LGO{s}~\rChiRi\ and the Jacobian ring structure of the cohomology as
obtained by the Koszul computation~\refs{\rMatter, \rCQG, \rSL}.

The purpose of this article is twofold. (a)~The general formula for
the Yukawa couplings in the Koszul computation~\rMatter\ is herein
shown to agree with the standard \LGO\ analysis. (b)~Some
computational short-cuts and identities are discussed which facilitate
toggling between the Koszul and the \LGO\ description. In addition, a
preliminary comparison is discussed with those exactly soluble
super\CFTs~\rGepnerIII\ where both a Koszul and a \LGO\ description
are also known, suggesting an equivalence of these three ring
structures.

The underlying reason for the existence of such a relation is the
common notion of the {\it function ring}---in fact {\it algebra}\ft{A
{\it ring} is a commutative group with respect to addition, equipped
with a multiplication that obeys the usual distributive laws.
Actually, most of the commonly encountered rings in physics here are
in fact {\it algebras}: the ring elements may be multiplied freely by
complex scalars, which form the {\it ground ring}, moreover the {\it
ground $\2{\hbox{\it field}}$}, $\IC$.}.
It is practically a tautology that a (quantum) field theory is
specified by its algebra of observables, that is, an additive group of
operators with the ring (algebra) structure specified through the
correlation functions; no Lagrangian or action functionals need be
known. Similarly, the ultimate point of view of algebraic geometry is
that a variety itself may be ignored and that `all one needs to know'
is the function ring~\rHartshorne. In \LGO{s} alike, the
superpotential serves for computations, but the desired information
lies in the Chiral ring.

Besides these general arguments, sample computations are also
included supporting the above claims and to demonstrate that the
{\it normalization} of the Chiral (Jacobian) ring elements remains
free to adjust for agreement with exactly soluble super\CFT\ models.
\ping

Roughly, our strategy is to establish a 1--1 correspondence between
the Koszul~\refs{\rBeast, \rMichael, \rMatter}, the \LGO~\rLGO,
working ultimately towards the Gepner-type~\refs{\rYukEx, \rGepnerIII,
\rGepner} computations. Since the last of these is most exact but
least general, the idea is to ``analytically continue'' (in moduli
space) these exact results of Gepner-type models to all Koszul
computations and so extend the work reported in Ref.~\rCQG.
Following Refs.~\refs{\rPhilip, \rMatter, \rCQG} and others, a few
examples will be presented in detail hoping that the general case
and limitations will be clear.

The article is organized as follows. In section~2, we describe the
typical results of the Koszul and the \LGO\ calculations, that is,
their respective descriptions of the spectrum of massless fields. In
section~3, we focus on the $(c,c)$-sector only and study a simple
model and its `ineffectively split' version. The $(a,c)$-sector is
then studied, in section~4, by applying our $(c,c)$-sector results to
the mirror of a well known model which was used to obtain a
3-generation theory~\rRolf. Section~5 contains a brief discussion of a
comparison with exactly soluble models and some additional comments.
Appendix~A presents a gauge theoretic introduction to the Koszul
computation while the equality of the Koszul ideal and the ideal of
the Chiral Ring structure is proven in  Appendix~B.

\newsec{Koszul {\it vs}.\ Landau-Ginzburg Orbifold Spectra}\noindent
At the risk of overlapping with the recent literature, a telegraphic
inventory of the standard toolboxes for algebraic geometry and for
(2,2)-supersymmetric effective field theory is presented here.
Refs.~\refs{\rBeast, \rMichael, \rMatter} and \rLGO, respectively, are
recommended for further information and details.

\subsec{Koszul field representatives}\noindent
The quickest (and perhaps conceptually simplest) way of introducing
this framework begins by starting with the embedding spaces of our
choice, which are products of flag varieties $\IF_{\vec{n}} \define
U(N)/( \prod_i U\6(n_i))$. The Bott-Borel-Weil theorem ensures
that (homogeneous, holomorphic) bundles over such spaces are
classified as representations of $\prod_i U(n_i)$, while cohomology
valued in these bundles furnishes representations of $U(N)$, where
$N= \sum_i n_i$. Representations of $SU(n)$ groups are rather well
known to physicists and suffice it here to note that $U(n) \approx
U(1) \times SU(n)$, so that representations of $U(n)$ are specified
as those of $SU(n)$, with an additional $U(1)$ charge. Appendix~A
provides more details about the application of the Bott-Borel-Weil
theorem.

As is well known, $SU(n)$ representations are easily assigned tensorial
variables. Irreducible $U(n)$-tensors have their indices
(anti)symmetrized and all possible traces are subtracted. Note that
traces are taken with the {\it only} invariant tensor of $U(n)$: the
Kronecker $\d_a^b$ symbol; the totally antisymmetric symbol
$\e^{a_1\ldots a_n}$ is an $SU(n)$, but {\it not} a $U(n)$
invariant---its $U(1)$-charge is $n$. \ping

Now, the spaces under study are not the entire flag spaces, but their
subspaces defined as the zero-set of a system of holomorphic
homogeneous polynomial constraint equations---complete intersections.
Each one of these equations may be written as $f(x)=0$, where $f(x)$
is the defining polynomial and may be written as
\eqn\eXXX{ f(x)= f_{(ab\cdots c)} x_a x_b \ldots x_c~,\qquad
           \deg(f)~ = ~d~,}
in terms of suitable homogeneous coordinates $x_a$.
This is equally well represented by $f_{(ab\cdots c)}$, which
we call the defining tensor (coefficient).

\bigskip
\vbox{\narrower\noindent\ninepoint\baselineskip=9pt plus1pt minus1pt
{\bf Notation:}\ Tensor coefficient will be indexed according to the
standard convention: subscripts for co-variant and superscripts for
contra-variant vectors. {\it However}, \,with a little forethought,
coordinates are indexed by subscripts although they are actually
contra-variant. Thus, $\f_a$ and $\g^a$ denote tensor coefficients
of a co- and contra-variant vector, so that $\f_a\g^a$ is invariant;
owing to this exceptional indexing of coordinates, $\f_a$ can be
contracted with $x_a$, but $\g^a$ and $x_a$ cannot; they transform
alike.}
\bigskip

It then follows that all tensor--valued cohomology on a complete
intersection submanifold is representable in terms of $U(N)$-tenors,
which are however further reduced by taking traces with the (dual)
defining tensors~\rBeast.

In other words, the tensor algebra on a flag space is generated by the
variously symmetrized traceless tensors, their usual products and
contractions with the aid of the Kronecker $\d_a^b$. However, on the
subspace ${\cal M}$ where $f(x) = f_{a\cdots c} x_a\ldots x_c$
vanishes, contractions with the aid of $f_{a\cdots c}$ are also
invariant and so tensors on $\cal M$ are irreducible only upon also
subtracting $f_{a\cdots c}$-traces.

In this approach, therefore, the required field representatives are
obtained as various (components of) tensor coefficients.
Typically\ft{This is so for complete intersections of hypersurfaces.
More general tensors arise for intersections of higher codimension
subspaces, i.e. when more constraint equations are required globally
than is the difference between the dimension of the embedding and the
embedded space; see Ref.~\rBeast.}, these will be totally symmetric
tensors or tensors which factorize into products of the totally
antisymmetric symbol $\e^{a_1\cdots a_n}$ and totally symmetric
tensors.

\subsec{Landau-Ginzburg field representatives}\noindent A \LGO\ model
is given in terms of a set of chiral superfields $X_a$ and a
superpotential $P(X)$; the kinetic term is typically taken to be the
flat one, $\int\rd^2\q\rd^2\ba\q\| X \|^2$ and may be ignored for most
part of the analysis.  Corresponding to projective spaces in the
(quasi)homogeneous flag-spaces, the superpotential $P(X)$ is
homogeneous. Such field theories fall in the general category of
Wess-Zumino models, which have been studied extensively in the last
two decades. A number of simple results can be derived about the
correlation functions, such as the fact that the chiral $n$-point
functions $\V{\prod_i(X_{a_i}\6(z_i))^{k_i}}$ do not depend on the
positions $z_i$ and are largely determined by the PCAC
theorem~\rCeGiPa.

Refs.~\rLGO\ provide a well-adopted technique for listing the states
in the $(c,c)$- and $(a,c)$-rings of a \LGO. Firstly, by
(quasi)homogeneity of the superpotential,
\eqn\eXXX{ \l P(X_a, Y_\a)~~ = ~~P(\l^{q_a}X_a, \l^{q_\a}Y_\a)~, }
with $q_a = n_a/d$ and $q_\a = n_\a/d$, where $d$, the total
degree, and $n_a,n_\a$ are integers. One then considers the general
cyclic symmetry $\Q$ with a diagonal action,
$X_a, Y_\a \mapsto (e^{2i\p\q_a}X_a), (e^{2i\p\q_\a}Y_\a)$.
When $\q_a=q_a$ and $\q_\a=q_\a$, $\Q$ is the $U(1)$-current $J_0$.
Clearly, as charges of a cyclic symmetry, $\q_a$ and $\q_\a$ are
defined only up to integers.
Upon passing to the quotient by $\Q$, the Hilbert space of the field
theory decomposes into sectors, one for each element of $\Q$. As each
element of the cyclic group $\Q$ is obtained as a power (denoted
$\ell$) of a chosen generator, this conveniently labels the sectors.
The general formula for the charges of the Ramond vacuum in the
$\ell$-twisted sector is~\rLGO~:
\eqn\eRCh{
  {J_0\atop\Jb_0} \Big|0\Big\rangle^{\ell}_R~ = \bigg\{\,
   \pm\Big[\sum\limits_{\Q_i(\ell) \not\in \ZZ}
               (\Q_i(\ell) - [\Q_i(\ell)] - {1\over2})\Big]~ +
     ~\Big[\sum\limits_{\Q_i(\ell) \in \ZZ}
               (q_i - {1\over2})\Big] \,\bigg\}
                                  \Big|0\Big\rangle^{\ell}_R~,
}
where $\Q_i(\ell)$, typically $\Q_i(\ell) = \ell\,\q_i$, is the
twisting angle of the $i^{th}$ field in the $\ell^{th}$ sector.  The
matching $\big|0\big\rangle^{\ell}_{(c,c)}$ and
$\big|0\big\rangle^{\ell+1}_{(a,c)}$ vacua are obtained by spectral
flows ${\cal U}_{(1/2,1/2)}$ and ${\cal U}_{(-1/2,1/2)}$, of charges
$({3\over2},{3\over2})$ and $(-{3\over2},{3\over2})$, respectively;
note the $\ell\to\ell+1$ shift in the ${\rm Ramond}\to(a,c)$ flow.

Upon obtaining a full list of $(c,c)$- and $(a,c)$-vacua for each
sector, one looks for charge-$(1,1)$ and charge-$({-}1,1)$ states
of the form $\F\ket{0}^{\ell}$ where  $\F\ket{0}^{\ell}$ is left
invariant by the quotient group $\Q$. Here, $\F$ is a polynomial
(possibly just the identity) in those fields which are left invariant
by the $\Q_i(\ell)$ action and $\ket{0}^{\ell}$ is the $\ell$-twisted
vacuum. Clearly, the charges of the polynomial $\F$ and the vacuum
$\ket{0}^{\ell}$ ought to add up to $(\pm1,1)$ for a marginal
operator (which corresponds to a massless field in the spacetime
effective field theory). Finally, two polynomials are considered
equivalent if they differ by a multiple of a gradient of the
superpotential. This is because, modulo the equations of motion, $\vd
P/\vd X$ is proportional to $\Db^2\Xb$, which is a descendant field
and yields no new information.

In addition, from general \CFT\ consideration, we know that (in the
\CFT\ limit) each twisted vacuum may be obtained from the untwisted
vacuum by multiplication with an appropriate twist-field.

\subsec{Koszul {\it vs.} Landau-Ginzburg dictionary}\noindent
By straightforward contraction, tensors are dual to formal
polynomials in $x_a$, $\rd x_a$ ($\rd x_a$ is replaced by spinors
$\j^a$ in a corresponding field theory) and their formal duals.
Corresponding to the fact that irreducible tensors on the hypersurface
$f(x)=0$ have no $f(x)$-traces, these formal polynomials are taken
modulo multiples of $f(x)$.

In fact, $U(N)$-linear transformations of the global
homogeneous coordinates on the embedding flag space cannot have any
intrinsic effect on the flag space or any of its manifolds. Since
\eqn\eXXX{ x_i \l_i{}^j \vd_j\, f(x)~  =  ~\l_{(i|}{}^{a}
              f_{a|b\cdots c)} x_i x_b \ldots x_c~, }
is the $U(N)$-linear transformation of $f(x)$, irreducible symmetric
tensors (dual to polynomials in $x_a$) on the hypersurface $f(x)=0$
vanish upon contraction with all but one index of $f_{(a\cdots c)}$.

Thus, we naturally expect the Koszul and the \LGO\ field
representatives to be each other's duals. Indeed, in the untwisted
sector of any \LGO\, this duality checks out readily. All \LGO\ field
representatives there are always polynomials of the same degree as
the superpotential; they produce polynomial deformations of the
superpotential:
\eqn\eXXX{
  \{ \f_{(ab \cdots c)} / \l_{(a|}{}^i f_{i|b\cdots c)} \}
 ~~\buildrel*\over\sim~~
    \{ \f(x) / x_a\l_a{}^i \vd_i\, f(x) \}~.}
The latter, in turn, always occurs as a subsector of
result of the Koszul computation, a subsector which is represented by
totally symmetric tensors precisely dual to the polynomial
deformations~\refs{\rBeast, \rMatter, \rCQG}.

Ref.~\rCQG\ has verified that this dual 1--1 correspondence goes
beyond the untwisted sector: all the Koszul representatives do have a
1--1 counterpart among the complete set of {\it untwisted and
twisted} states. Moreover, just as the \LGO\ representatives are of
the form $\F(X)\ket{{\rm vacuum}}^{\ell}$, the Koszul
representatives factor into a totally symmetric tensor, which is dual
to $\F(X)$, and a (product of) totally antisymmetric symbol(s) which
correspond (at least formally) to the twist fields. This
correspondence certainly covers the obvious properties such as the
various charges. To further this relation, we will propose a
``polynomial'' equivalent of the antisymmetric symbol $\e^{a_1\ldots
a_n}$ at least in the sense that it fits the Koszul {\it vs.} \LGO\
correspondence.\ping

Beyond the general discussion presented above, it is a different (and
rather more involved) issue to determine precisely which tensors are
required to span completely a given cohomology group on ${\cal M}$.
In \CY\ compactifications of heterotic superstrings, elements of
$H^1(\TM)$ and $H^1(\CM)$ correspond, respectively, to the
charge-$(1,1)$ and charge-$({-}1,1)$ states of the corresponding \LGO,
and we will use the classical geometry and \LGO\ notation
interchangeably. For \CY\ complete intersections in products of
projective spaces, $H^1(\TM)$ is not infrequently well described by
deformations of the defining polynomials~\rPhilip. In general (product
of any flag-spaces and ${\cal M}$ not necessarily \CY), the fully
fledged Koszul computation~\refs{\rBeast, \rMichael, \rMatter} is
necessary to compute $H^1({\cal T})$; $H^1(\CM)$ is also obtained by
the same method. We now focus on these two sectors in turn.

\newsec{The $(c,c)$-Ring}\noindent
Rather than attempt to directly deal with the general case and
unnecessarily clutter the notation, we consider a simple example where
the $(c,c)$-sector contains both twisted and untwisted states;
computational details are included in order to facilitate future
applications of the present results to other models.

Before doing so, however, a remark is in order. As noted below, we
shall use a trick to extend the Yukawa coupling formula of
Ref.~\rPhilip\ to the twisted $(c,c)$-states. Undeniably, this will
be possible only in a subclass of models. However, the obtained
formula agrees in all detail with the general Koszul
computation~\refs{\rBeast, \rMatter}. Since the correspondence is
equally general~\rCQG, we conclude that the extension of the Yukawa
coupling will be valid even when the calculational trick employed
below will not be available. On the other hand, this also proves that
the Koszul calculations of Yukawa couplings~\refs{\rBeast, \rMatter}
in fact are as exact as those obtainable in \LGO{s}, when those are
available.

\subsec{A sample model}
Consider the \CY\ manifold of the type
\eqn\eQDf{
{\cal Q} \in \K{4\cr1\cr}{4&1\cr0&2\cr}^{2,86}_{-168}~~:~~
 \cases{f(x)   = f_{abcd}\, x_a\, x_b\, x_c\, x_d &= 0~,\cr
        g(x,y) = g_{a\,\a\b}\, x_a\, y_\a\, y_\b  &= 0~,\cr}}
where $(x_0,\ldots,x_4)$ and $(y_0,y_1)$ are homogeneous
coordinates on $\CP4 \times \CP1$, denoted by the `bra' part,
$\Big[{4\atop1}\Big|$.  Obviously, $f_{abcd}$ is totally symmetric in
all its indices, while $g_{a\,\a\b} = g_{a\,\b\a}$.  The two
columns in the `ket' part of the matrices above denote the
homogeneity of $f(x)$ and $g(x,y)$. The matrix represents an entire
family of complex manifolds, of the same topological type but with
the complex structure parametrized by the various choices of the
coefficients in $f(x)$ and $g(x,y)$ and $\cal Q$ is a member of this
family.  Generic\ft{``Generic'' means ``all except a subset of
strictly smaller dimension''.} choices of $f(x)$ and $g(x,y)$ produce
smooth $\cal Q$ and the subscripts on the matrices above denote the
Euler characteristic while the superscripts denote $b_{1,1}$ and
$b_{2,1}$ of such a generic manifold.
\ping

The \LGO\ is obtained by assigning chiral superfields $X_a$ and
$Y_\a$ to the coordinates $x_a$ and $y_\a$ and using $P(X,Y) \define
f(X) + g(X,Y)$ as the superpotential~~\rBVW. For a generic choice of
$f(X)$ and $g(X,Y)$, the superpotential is non-degenerate. We easily
find:
\eqn\eQCh{ {\cal Q} ~:~
 \q_X=q_X=\frc14~, \quad \q_Y=q_Y=\frc38~, \qquad \Q = \ZZ_8~. }
The formula~\eRCh\ now produces the charges of all eight Ramond
ground states and spectral flow then provides the Neveu-Schwarz
vacua. For completeness, Table~1 lists the five classes of vacua.

\subsec{Field representatives for $\cal Q$}\noindent
The Koszul computation provides a full complement~:
\eqn\eQT{ \eqalign{
 H^1({\cal Q},{\cal T_Q}) ~~\sim~~
 & \big[\,\{\, \f_{(abcd)} \,\}/
     \{\, f_{e(abc}\, \l_{d)}{}^e \,\}\,\big]_{45}~,        \cr
 \noalign{\vglue1mm}
 &~~\oplus ~\big[\,\{\, \vf_{a(\b\g)} \,\}/
     \{\, g_{d\,\b\g}\, \l_a{}^d \oplus
          g_{a\,\d(\b}\, \l_{\g)}{}^\d \,\}\,\big]_{11}~,   \cr
 \noalign{\vglue1mm}
 &~~\oplus ~\big[\,\{\, \e^{\a\b}\,\vq_{(abc)} \,\}/
     \{\, \e^{\a\b}\, f_{abcd}\, \l^d \,\} \,\big]_{30}~,  \cr}}
of 86 independent tensor components, as counted by the subscripts;
$a,b..=1,\ldots,5$ and $\a,\b..=1,2$.  Here $\l_a{}^b$, $\l_\a{}^\b$
and $\l^a$ are reparametrization degrees of freedom which can be used
to ``gauge away'' 25, 4 and 5 components, respectively, of the $\f$,
$\vf$ and $\vq$. Note that the $\l_a{}^b$-``gauge equivalence'' occurs
in two places; the above subscripts imply that all of it was used up
to gauge away components of $\f_{(abcd)}$, rather then of
$\vf_{a(\b\g)}$. Clearly, this choice is subject to the practitioner's
whim, provides equivalent representatives for the elements of
$H^1({\cal Q},{\cal T_Q})$ and---most importantly---relates many
Yukawa couplings.

Upon contracting the first 56 tensors with appropriate coordinates,
the Reader will promptly recognize the deformations of the defining
polynomials $f(x)$ and $g(x,y)$, taken modulo linear
reparametrizations~\rPhilip. The last 30 elements however do not have
such a simple interpretation and cannot be found by polynomial
deformations~\rPDM. Note the anti-symmetric symbol $\e^{\a\b}$
occurring in these.
\ping

A quick glance at Table~1 verifies that only the $(c,c)$-vacua with
$\ell=0,4$ may be used to construct charge-$(1,1)$ states.

For $\ell=0$, all $X_a, Y_\a$ fields are left invariant and since the
vacuum charges are zero, we need polynomials of charge $(1,1)$. In
view of Eq.~\eQCh, the untwisted $(c,c)$-sector is spanned by
polynomials of bi-degree (4,0) and (1,2) in $X_a,Y_\a$---precisely
the polynomial deformations of $f(X)$ and $g(X,Y)$ and {\it dual} to
the first 56 representatives in~\eQT. The number of independent
such polynomials, taken modulo the ideal $\Im[\vd P]$, is
found~\rLGO\ as the coefficient of the $t^8$-term in
\eqn\eXXX{ {\rm P}(t^8)~
 = ~\Big( {1-t^{8(1-q_X)} \over 1-t^{8q_X}} \Big)^5
      \Big( {1-t^{8(1-q_Y)} \over 1-t^{8q_Y}} \Big)^2~
 = ~ 1 + \ldots + 56\,t^8 + \ldots }
It is not a least bit obvious that the ideal generated by the
gradients of $P(X,Y)=f(X)+g(X,Y)$ is the same as the one dictated by
the Koszul computation, but this is nevertheless true~\rCQG.

Now, for $\ell=4$, only the $X_a$ fields are left invariant and since
the vacuum charges are both $\inv4$, we need polynomials of charge
$(\frc34,\frc34)$. The twisted $(c,c)$-sector is therefore spanned by
cubic polynomials in $X_a$---corresponding to the last 30
representatives in~\eQT. Here, the \LGO\ and the Koszul ideal are
more easily shown to be the same. What remains is to identify, at
least formally, the anti-symmetric symbol $\e^{\a\b}$ with the twist
field which creates $\ket{\inv4,\inv4}^4$ from $\ket{0,0}^0$. The
calculation of the Yukawa coupling (below) will support this
correspondence in rather more detail.

Suffice it here to note that $\e^{\a\b}$ transforms the same as
$(y\rd y)$, which has charge $(\frc34,\frc34)$ and is therefore dual
to $\ket{\inv4,\inv4}^4$ just as $\q_{abc}$ is dual to the monomial
$x_a\,x_b\,x_c$. A related fact is that all tensorial
representatives~\eQT\ have an integral charge under the $\ZZ_8$
symmetry~\eQCh.\ping

It is rather well known that the polynomial deformation method for
computing the ${\bf27}^3$ Yukawa couplings~\rPhilip\ applies {\it
verbatim} to the untwisted $(c,c)$-sector of \LGO{s}. Refs.~\rMatter\
shows how to (a)~translate this method to the Koszul computation and
(b)~extend to all of $H^1({\cal T})$, whether corresponding to
twisted or untwisted $(c,c)$-states. The resulting ring structure
will here be called `Jacobian', borrowing from the standard case
of a single hypersurface (see Refs.~\rSL\ and references therein).

Given the above 1--1 correspondence between Koszul and \LGO\ field
representatives, it is of course a simple matter to translate this
result into the \LGO\ formalism. However, we presently re-derive it
from the \LGO\ framework itself, using a little trick for which we
need a model which is closely related to $\cal Q$.

\subsec{The `ineffectively split' model}\noindent
For many complete intersection manifolds, there is a sequence of
related manifolds obtained by so called `ineffective splitting', first
discussed and named\ft{Given that this technique is quite effective,
this appears to be somewhat of a misnomer.} in the third article in
Refs.~\rCYCI; see also Refs.~\refs{\rBeast, \rElusive}.

To be precise, {\it every} complete intersection can be split, but
this may or may not change the manifold; a `split' is called
`ineffective' when the manifold does not change although the
embedding is changed through `splitting'. If the complete
intersection can be written as a hypersurface in a product of a
complex 1-dimensional and a complex 3-dimensional space, the simple
form of `ineffective splitting' as described below is possible. More
general forms of `splitting', by introducing more than two new
variables, will provide `ineffective splits' of many other examples
(see for example p.~279 of Ref.~\rBeast). To the best of our
understanding, the precise limitation of this process is not known in
general.

Consider the following `ineffectively split' version of $\cal Q$:
\eqn\eXXX{
 \Tw{\cal Q}
  \in \K{4\cr1\cr1\cr}{4&1&0\cr0&0&2\cr0&1&1\cr}^{2,86}_{-168}~~:~~
   \cases{f(x)   = f_{abcd}\, x_a\, x_b\, x_c\, x_d &= 0~,\cr
          m(x,z) = m_{a\,B}\, x_a\, z_B             &= 0~,\cr
          h(y,z) = h_{\a\b\,C}\, y_\a\, y_\b\, z_C  &= 0~.\cr}}
The relation between $\cal Q$ and $\Tw{\cal Q}$ is given by $g(x,y) =
\det\big({\vd(m,h)\over\vd(z_1,z_2)}\big)$. Since the topological
numbers remained the same, $\Tw{{\cal Q}}$ and $\cal Q$ are identical,
except that not all polynomial deformations of the complex structure
in one family can be attained by polynomial deformations in the other
and {\it vice versa}.

Without much ado, the Koszul computation produces
\eqn\eQTw{ \eqalign{
 H^1(\Tw{\cal Q},{\cal T}_{\Tw{\cal Q}}) ~~\sim~~
 & \big[\,\{\, \f_{(abcd)} \,\}/
     \{\, f_{e(abc}\, \l_{d)}{}^e \,\}\,\big]_{45}~,        \cr
 \noalign{\vglue1mm}
 &~~\oplus ~\big[\,\{\, \m_{a\,B} \,\}/
     \{\, m_{c\,B}\, \l_a{}^c \oplus
          m_{a\,C}\, \l_B{}^C \,\}\,\big]_6~,               \cr
 \noalign{\vglue1mm}
 &~~\oplus ~\big[\,\{\, \h_{(\a\b)C} \,\}/
     \{\, h_{\d(\a|C}\, \l_{|\b)}{}^\d \oplus
          h_{\a\b\,D}\, \l_C{}^D \,\}\,\big]_2~,            \cr
 \noalign{\vglue1mm}
 &~~\oplus ~\big[\,\{\, \e^{\a\b}\e^{AB}\,\vq_{(abc)} \,\}/
 \{\, \e^{\a\b}\e^{AB}\, f_{abcd}\, \l^d \,\} \,\big]_{30}~,\cr
 \noalign{\vglue1mm}
 &~~\oplus ~\big[\,\{\, \e^{\a\b}\,\vf_c \,\}/
     \{\, \e^{\a\b}\,m_{c\,B}\, \l^B \,\}\,\big]_5~.        \cr}}

Similarly to $\cal Q$, the $\Tw{\cal Q}$ \LGO\ is constructed with
the superpotential $P(X,Y,Z) = f(X) + m(X,Z) + h(Y,Z)$, charges
\eqn\eQSC{ \ha{\cal Q} ~:~ \q_X=q_X=\inv4~,
 \quad \q_Y=q_Y=\inv8~, \quad \q_Z=q_Z=\frc34~, \qquad \Q = \ZZ_8~, }
and the vacua listed in Table~2. Again, the charge-$(1,1)$
states are found amongst the untwisted and $\ell=4$ twisted states.
This time, however, the $\ell=4$ twisted vacuum has both charges
equal to $\frc34$, whence the charge-$(1,1)$ state is obtained by
multiplying this only by linear combinations of $X_a$. Of course,
owing to more variables, there are now more untwisted states, so the
net count is the same---86.

To simplify the presentation and also for later convenience, we now
shift to the special case, where
\eqn\eFM{ f_{abcd}~    = ~\cases{ 1 & $a,b,c,d$ all equal,\cr
                                  0 & otherwise,          \cr}
  \qquad  g_{a\,\b\g}~ = ~\cases{ 1 & $a,\b,\g$ all equal,\cr
                                  0 & otherwise,          \cr}}
for $\cal Q$, and
\eqn\eFMS{ {\eqalign{
           f_{abcd}~    = ~\cases{ 1 & $a,b,c,d$ all equal,\cr
                                   0 & otherwise,          \cr}
 &\qquad   m_{a\,B}~    = ~\cases{ {-}\e_{aB} & $a,B=1,2$, \cr
                                   0 & otherwise,          \cr} \cr
  \qquad   h_{\a\b\,C}~ =~&\cases{ 1 & $\a,\b,C$ all equal,\cr
                                   0 & otherwise,          \cr} \cr}}}
for $\Tw{\cal Q}$. The respective \LGO\ superpotentials become
\eqn\eQP{ P_{\cal Q}(X,Y) ~~=~~
  \sum_{\a=1}^2\big( X_\a^4 + X_\a\, Y_\a^2\big)~ +
    ~X_3^4 + X_4^4 + X_5^4~,
}
and (upon a convenient rescaling $Z_1\to Z_2$, $Z_2\to{-}Z_1$)
\eqn\eXXX{ P_{\Tw{\cal Q}}(X,Y,Z) ~~=~~
  \sum_{A=1}^2\big( X_A^4 + X_A\,Z_A + Y_A^2\,Z_A\big)~ +
    ~X_3^4 + X_4^4 + X_5^4~.
}
It is of course convenient to use a monomial basis for the states and
Table~3 presents a comparative list of the 86 field representatives
of $\cal Q$, $\Tw{\cal Q}$, both the Koszul and the \LGO\ version.

First of all, all Koszul representatives have integral charges under
the $\ZZ_8$ (are left invariant by it); also, the total scaling charge
of the holomorphic volume form $(x\rd^4x)(y\rd y)(z\rd z)$ is 3.
Now, the Koszul representatives which correspond to twisted states in
fact have zero scaling charge, while those corresponding to untwisted
states have non-zero scaling charges, equal here to $\pm1$.
We will see below that this is also true of the $(a,c)$-sector, taken
modulo the charge of the full (holomorphic) volume-form.

Next, notice that the Koszul representative
$\e^{\a\b}\e^{AB}\vq_{(abc)}$ in
$H^1(\Tw{\cal Q},{\cal T}_{\Tw{\cal Q}})$ corresponds to an untwisted
state although it contains antisymmetric symbols. Since
\eqn\eXXX{ X_a X_b X_c Y_\a Y_\b \ket{0,0}^0
           ~~\buildrel*\over\sim~~
           \e^{\a\b}\e^{AB}\vq_{(abc)}~,}
we conclude that $\e^{\a\b}\e^{AB}$ ought to be dual to $Y_\a Y_\b$.
Indeed, this is not hard to prove, since these two quantities can be
contracted using the defining tensors:
\eqn\eEpsYY{ \e^{\b\d}\e^{AB}~h_{(\a\b)A} h_{(\g\d)B}~Y_\a Y_\g }
is invariant; recall that coordinates are indexed by sub-scripts
although they are contra-variant. Thus, $\e^{\b\d}\e^{AB}$
is dual to $Y_\a Y_\g$ $(= Y_1 Y_2)$, much like $\vq_{(abc)}$ is dual
to $X_a X_b X_c$.

It only remains to interpret the single $\e$'s. To this end we
consider the Yukawa couplings in the $(c,c)$-sector.

\subsec{Yukawa couplings}\noindent
The 86 field representatives in Table~3 are easily divided into three
groups: (1)~those which are untwisted in both ${\cal Q}$ and $\Tw{\cal
Q}$, denoted {\bf U}, (2)~those which are untwisted in ${\cal Q}$ but
twisted in $\Tw{\cal Q}$, denoted {\bf V}, and (3)~those which are
twisted in ${\cal Q}$ but untwisted in $\Tw{\cal Q}$, denoted {\bf W}.

The Yukawa coupling is a scalar product of the type
\eqn\eXXX{ \k_{ijk} ~~=~~ \bra{0,0}
            \ket{{-}3,0}^1 \ket{0,{-}3}^7 \ket{1,1}^{\ell_i}_i
             \ket{1,1}^{\ell_j}_j \ket{1,1}^{\ell_k}_k~, }
where $\ket{1,1}^{\ell_i}_i$ are the charge-$(1,1)$ states (\LGO\
field representatives) from Table~3 and  of course
\eqn\eTcc{ \ell_i+\ell_j+\ell_k ~\id~ 0 \pmod{d}~, }
with $d=8$ here, is the twist-number selection rule in the
$(c,c)$-sector, and the $J_0,\Jb_0$-charges have been balanced
already.
Note that
the $\ZZ_8$ selection rule~\eTcc, $\ell_i+\ell_j+\ell_k \id 0 \pmod8$,
applied both for $\cal Q$ and $\Tw{\cal Q}$, guarantees that
\eqn\eXXX{ \V{{\bf U}, {\bf U}, {\bf U}}~,\quad
           \V{{\bf U}, {\bf V}, {\bf V}}~,\quad
           \V{{\bf U}, {\bf W}, {\bf W}} }
are the only non-zero types of Yukawa couplings.\ping

The first of these types is entirely in the untwisted $(c,c)$-sector
of both $\cal Q$ and $\Tw{\cal Q}$. Therefore, the polynomial
deformation formula of Ref.~\rPhilip\ applies and for a concrete value
we only need to calculate $Q \define \det\big[ \vd^2 P \big]$,
the `top' element of the Chiral (Jacobian) ring. Straightforwardly,
we have
\eqna\eQQ
$$ \eqalignno{
 Q_{\cal Q}
   &\simeq ~2^{12} 3^2 5^2\, X_1^3\,X_2^3\,X_3^2\,X_4^2\,X_5^2~,
                                                     &\eQQ{a}\cr
 Q_{\Tw{\cal Q}}
   &\simeq ~2^{12} 3^2 7^2\, X_1^3\,X_2^3\,X_3^2\,X_4^2\,X_5^2~.
                                                     &\eQQ{b}\cr}
$$
each of which has to be understood as representative of an
equivalence class, up to the ideal generated by the partials
\eqn\eId{ \Im[\vd P_{\cal Q}] ~\sim~
 \bigoplus_{\a=1,2}\!\big\{ (4X_\a^3+Y_\a^2), ~(2X_\a Y_\a) \big\}
  \oplus \bigoplus_{a=3,4,5}\!\!\big\{ 4X_a^3 \big\}~, }
and
\eqn\eIdS{ \Im[\vd P_{\Tw{\cal Q}}] ~\sim~
 \bigoplus_{\a=1,2}\!
  \big\{ (4X_A^3+Z_A), ~(2Y_A Z_A), ~(X_A+Y_A^2) \big\}
  \oplus \bigoplus_{a=3,4,5}\!\!\big\{ 4X_a^3 \big\}~, }
respectively. The overall numerical factors $2^{12} 3^2 5^2$ and
$2^{12} 3^2 7^2$, respectively, in Eqs.~\eQQ{a} and~\eQQ{b} are easily
absorbed in an overall rescaling of the superpotential and will be
ignored hereafter. Thus up to wave function renormalization,
the Yukawa coupling is given by
\eqn\eYukU{
\k_{uuu} ~~=~~ {U^3\over Q_{\cal Q}}
}
and similarly for $\Tw{\cal Q}$.
\ping

The remaining $\V{{\bf U}, {\bf V}, {\bf V}}$-type
couplings are straightforward to evaluate in $\cal Q$, since both
{\bf U} and {\bf V} are untwisted there. Similarly, the
$\V{{\bf U}, {\bf W}, {\bf W}}$-type couplings are easily
evaluated in $\Tw{\cal Q}$. Once we know the values of these
couplings in the version where they are untwisted, we require that
the value of the coupling in the twisted version be the same. Thus,
for example,
\eqn\eXXX{ \bra{0,0} \ket{{-}3,0}^1 \ket{0,{-}3}^7
            \big(X_3 Y_1 Y_2 \ket{0,0}^0\big)^2
             \big(X_4^2 X_5^2 \ket{0,0}^0\big) }
is the untwisted version, in $\cal Q$, of the Yukawa coupling
\eqn\eXXX{ \bra{0,0} \ket{{-}3,0}^1 \ket{0,{-}3}^7
            \big(X_3 \ket{\frc34,\frc34}^4\big)^2
             \big(X_4^2 X_5^2 \ket{0,0}^0\big) }
in the twisted sector of $\Tw{\cal Q}$. The former of these easily
produces $Q_{\cal Q}$ upon using the ideal~\eId: the Yukawa coupling
is 16 (up to the irrelevant overall numerical coefficient). The latter
evaluates to
\eqn\eXXX{ \big(X_3^2 X_4^2 X_5^2\big) \bra{0,0}
            \ket{{-}3,0}^1 \ket{0,{-}3}^7 \ket{\frc34,\frc34}^4
             \ket{\frc34,\frc34}^4 \ket{0,0}^0 }
Comparing with Eq.~\eQQ{b}, it must be that
\eqn\eTSS{
 \Big[\, \ket{\frc34,\frc34}^4 \,\Big]^2 ~~=~~
  16\big(X_1^3\,X_2^3\big)\ket{0,0}^0 ~\simeq~
   \big(Z_1\,Z_2\big)\ket{0,0}^0~, }
in $\Tw{\cal Q}$, at least in the sense that this fits the Yukawa
couplings and hence the ring structure. A similar comparison of the
$\V{{\bf U}, {\bf W}, {\bf W}}$-type couplings, we derive
that, in $\cal Q$,
\eqn\eTS{ \Big[\, \ket{\inv4,\inv4}^4 \,\Big]^2
            ~~=~~ \big(X_1\,X_2\big)\ket{0,0}^0~.}

Formally, we can also set
\eqn\eEpZZ{
 \ket{\frc34,\frc34}^4 ~=~ \sqrt{Z_1\,Z_2}\ket{0,0}^0~,\qquad
  {\rm~in~} \Tw{\cal Q}~, }
and
\eqn\eEpXX{
 \ket{\inv4,\inv4}^4 ~=~ \sqrt{X_1\,X_2}\ket{0,0}^0~,\qquad
  {\rm~in~} {\cal Q}~. }
Because of the square-root, the fields in the super\CFT\
corresponding to $\sqrt{Z_1Z_2}$ and $\sqrt{X_1X_2}$ must have a
branch-cut which is a well-known property of twist-fields.
This supports our identification of $\sqrt{Z_1Z_2}$ as the
twist-field that creates $\ket{\inv4,\inv4}^4$ from $\ket{0,0}^0$ in
$\Tw{\cal Q}$ and $\sqrt{X_1X_2}$ as the one that creates
$\ket{\frc34,\frc34}^4$ from $\ket{0,0}^0$ in $\cal Q$.
The $J_0,\Jb_0$-charges of $\sqrt{X_1X_2}$ and $\sqrt{Z_1Z_2}$ clearly
satisfy this identification. \ping

Looking back at the Koszul field representatives, the quadratic
relation~\eTSS\ implies that, on $\Tw{\cal Q}$, $\e^{\a\b}$
is dual to $\sqrt{Z_1 Z_2}$. Indeed, this is easy to prove, since the
product of the squares of these objects is easily contracted into a
scalar:
\eqn\eEpsZZ{
 \big[ \e^{\a\g} \e^{\b\d}~h_{\a\b\,A} h_{\g\d\,B}~Z_A Z_B \big]
  \quad\hbox{is invariant on}\quad \Tw{\cal Q}~. }
Similarly, on $\cal Q$, $\e^{\a\b}$ is dual to
$\sqrt{X_1 X_2}$; indeed, this again is easy to prove:
\eqn\eEpsXX{
 \big[ \e^{\a\g} \e^{\b\d}~g_{a\,\a\b} g_{b\,\g\d}~X_a X_b \big]
  \quad\hbox{is invariant on}\quad {\cal Q}~.}

Clearly, the same calculation can be repeated for any other
non-degenerate choice of the superpotential; the choice~\eFM\
and~\eFMS\ merely simplified the derivation. \ping

Thus, we have derived {\it the same} ``polynomial'' representative for
the twist fields
\item{(a)} by comparing the Yukawa couplings in the $(c,c)$-sector of
           a \LGO\ and its `ineffectively split' version, and
\item{(b)} by identifying the twist-fields with $\e$'s in the Koszul
           representative and contracting with the defining tensor
           coefficients. Note that such relations, Eqs.~\eEpsYY,
           \eEpZZ, \eEpXX, \eEpsZZ\ and \eEpsXX, hold precisely
           because of the way the tensor algebra on the submanifold
           was defined in Section~2.1.\vglue0mm\noindent
With such a ``polynomial'' representative, the Yukawa couplings can be
computed for all $(c,c)$-states, untwisted and twisted.

It is a straightforward matter to verify that the resulting
numerical values of the couplings are identical to those calculated
entirely within the Koszul computation framework~\refs{\rBeast,
\rMatter}. In view of the non-renormalization theorem of Ref.~\rDGNR,
this should not come as a surprise. This identifies the Koszul ring
structure with the Chiral ring structure---up to the normalization of
the field representatives. In other words, the foregoing discussion of
Yukawa couplings pertains to the so called ``un-normalized''
couplings, with the kinetic terms of the various fields left
undetermined. We will return to this later.

\subsec{The general case}
With the above we have shown that indeed the Koszul computation
agrees with the Landau-Ginzburg ring calculation. Unfortunately,  in a
number of cases we know of no non-degenerate \LGO\ corresponding to
the manifold. Yet, our technique may equally well be applied in those
situations.

Let ${\cal M}$ be a family of manifolds, each defined as a
hypersurface in a product of $N$ projective spaces
$\CP{n_1}\times\ldots\CP{n_N}$ where $x_j^{i}$ are the coordinates on
the $\CP{n_i}$. The hypersurface is given by $M$ constraints of the
form $f^{(m)}_{a\ldots b} x^{(i)}_a\ldots x^{(j)}_b=0$,
$m=1,\ldots,M$, $i,j=1,\ldots,N$.

The polynomial deformations will as usual be spanned by polynomials
taken modulo the Jacobian ideal. In order to find a representative for
the various $\e^{\ldots}$ we consider the square, $\e^2$ contracted
with  the appropriate number of the defining tensors,
$f_{\ldots}^{(m)}$ and coordinates, $x_j^{(i)}$, to make the
expression invariant. Just as in the above example, $\e^{\ldots}$ will
be the dual of the square root of the product of the corresponding
$x_j^{(i)}$'s. By using the Koszul computation and the above
identification, we can write explicit monomial representatives for all
complex deformations---although we do not know the corresponding
\LGO~.

With the $(c,c)$ ring at hand, we then go on to compute the Yukawa
couplings using the usual technique. In fact, the knowledge of the
ring structure may help us in finding the underlying conformal field
theory.

\newsec{The $(a,c)$-Ring}
Having extended the polynomial deformation formula for the Yukawa
couplings~\rPhilip\ to the complete $(c,c)$-sector of a \LGO\ model,
the extension of this formula to the $(a,c)$-sector is now easily
obtained by using mirror symmetry. Again, as a calculational trick,
we will use the `mirror map'.

Although the list of models for which the `mirror model' is known is
growing~\rGCM, it will not be known in the general case. As with the
$(c,c)$-ring, we will again show that the results of the Koszul
computation agree with those obtained by extending the Yukawa
coupling formula~\rPhilip\ with the aid of the mirror map. It is
important to realize at this point an inherent limitation of the
Koszul computation: while the choice of the complex structure is
controlled, the choice of the K\"ahler class is not, and so a generic
choice is implied. This means that certain parameters (and so Yukawa
couplings) will remain undetermined by the Koszul computation
although they can be calculated by other geometrical
means~\refs{\rBeast, \rCubi}. Our aim is to show that these can be
chosen so as to agree with the \LGO\ calculations. With this
limitation in mind, the Koszul-calculated Yukawa couplings will be
show to be equal to the \LGO\ results. We present two examples in
detail to facilitate the application to other models.

\subsec{An almost trivial example}
Let $\widehat{\cal Q}$ be the mirror model of $\cal Q$. Then, the
$(a,c)$-sector of $\cal Q$ maps 1--1 to the $(c,c)$-sector of
$\widehat{\cal Q}$. Thus, we employ the full $(c,c)$-extension of the
Yukawa coupling formula~\rPhilip\ to the $(c,c)$-sector of
$\widehat{\cal Q}$ and then map the results back to the $(a,c)$-sector
of $\cal Q$.

As seen from Table~1, there are only two charge-$({-}1,1)$ states in
the $(a,c)$-sector of $\cal Q$: two $(a,c)$-vacua in fact:
$\ket{{-}1,1}^3$ and $\ket{{-}1,1}^4$. The Koszul computation (and
also a straightforward application of the Lefschetz hyperplane
theorem) assures us that these must correspond to the pull-backs of
the K\"ahler forms of $\IP^4$ and $\IP^1$. Their (geometrical) Yukawa
couplings are easily calculated as they are isomorphic to generic
hyperplanes in $\IP^4$ and $\IP^1$, respectively~\rBeast. In the
matrix notation of~\eQDf, we have $J_x \approx \big|{1\atop0}\big]$
and $J_y \approx \big|{0\atop1}\big]$. The Yukawa couplings are then
simply the number of intersection points:
\eqna\eGGG
$$ \eqalignnotwo{
 \V{J_x,J_x,J_x} &=
 ~\EU\left.\left.\K{4\cr1\cr}{4&1\cr0&2\cr}
                  \right|\matrix{1&1&1\cr0&0&0\cr}\right]
                 &= 8~,                                &\eGGG{a}\cr
 \V{J_x,J_x,J_y} &=
 ~\EU\left.\left.\K{4\cr1\cr}{4&1\cr0&2\cr}
                  \right|\matrix{1&1&0\cr0&0&1\cr}\right]
                 &= 4~,                                &\eGGG{b}\cr
}
$$
all others being zero. These results are however subject to
world-sheet instanton corrections~\rDSWW, which are insignificant only
in the large size limit. That is, the corrections to the above results
come as a formal power series in powers of the smallest {\it
characteristic} size of the `internal' manifold, in units of Planck
length. This smallest characteristic size is simply the size of
minimal $\IP^1$'s embedded in the \CY\ manifold and is unrelated to
the overall size. For example, orbifolds have singular points, and the
minimal $\IP^1$'s are in fact these singular points themselves,
regardless of the overall size of the orbifold.

Clearly, therefore, the above (geometric) result is unreliable unless
an independent argument ensures that world-sheet instanton effects are
negligible. No such argument is available for \LGO{s}, and in fact one
expects that at least some of the characteristic sizes are small.\ping

However, our strategy is to relate the Yukawa couplings in the
$(a,c)$-sector of $\cal Q$ to those in the $(c,c)$-sector of the
mirror model, $\widehat{\cal Q}$. The latter couplings are protected
by the non-renormalization theorem of Ref.~\rDGNR, whereupon the
inferred Yukawa couplings in the $(a,c)$-sector of $\cal Q$ will also
be exact.

For the specific choice of superpotential~\eQP, the mirror model is
easily constructed~\rGCM. Note that the \LGO\ model $\cal Q$ consist
of five ``building blocks'', uncoupled up to the $\ZZ_8$ GSO-like
projection: two copies of a $D_5$-type model, $p_\a =
X_\a^4+X_\a Y_\a^2$, $\a=1,2$ and three copies of a $A_3$-type model,
$p_a = X_a^4$, $a=3,4,5$. The mirror of $A_k$-type models has the
same superpotential, but is a $\ZZ_{k+2}$ quotient thereof. One way
to construct the mirror of $D_k$-type models is to `transpose' the
superpotential and then divide by an extra $\ZZ_2$~\rGCM.
All in all, the mirror model of~\eQP\ will have the superpotential
\eqn\eQPM{ P_{\ha{\cal Q}}(\hat{X},\hat{Y}) ~=~
 \sum_{\a=1}^2 (\hat{X}_\a^4\hat{Y}_\a + \hat{Y}_\a^2)~
  ~+~ \hat{X}_3^4 + \hat{X}_4^4 + \hat{X}_5^4~, }
This dictates that
\eqn\eXXX{
 q(\hat{X}_\a)=\inv8~,\quad q(\hat{Y}_\a)=\inv2~,\quad \a=1,2~,
 \qquad q(\hat{X}_a)=\inv4~,\quad a=3,4,5~. }
With these variables, we have
\eqn\eXXX{ \vq_x ~\define~
            \hat{X}_1 \hat{X}_2 \hat{X}_3 \hat{X}_4 \hat{X}_5~,\qquad
           \vq_y~ \define ~\hat{Y}_1 \hat{Y}_2~, }
as the mirror images of $J_x$ and $J_y$; actually, the exact
identification as to which $\vq$ is the mirror of which $J$ will be
clear from the analysis below. The general results of Refs.~\rSL\ then
imply that these two representatives will be valid for a generic
choice of the superpotential $P_{\cal Q}$, not just the special
case~\eQP.

For the superpotential~\eQPM, the Jacobian ideal is generated by
\eqn\eIdM{ \Im[\vd P_{\ha{\cal Q}}] ~\sim~
 \bigoplus_{\a=1,2} \big\{ (4\hat{X}_\a^3\hat{Y}_\a),
                          ~(\hat{X}_\a^4+2\hat{Y}_\a) \big\}
  \oplus \bigoplus_{a=3,4,5}\big\{ 4\hat{X}_a^3 \big\}~, }
and
\eqn\eXXX{ Q_{\ha{\cal Q}} ~=~
 2^{10}3^3\,\hat{X}_1^6\hat{X}_2^6\hat{X}_3^2\hat{X}_4^2\hat{X}_5^2
 ~\simeq~
 2^{12}3^3\,\hat{X}_1^2\hat{Y}_1\hat{X}_2^2\hat{Y}_2
             \hat{X}_3^2\hat{X}_4^2\hat{X}_5^2~. }
It is easy to show that $\vq_x^{~3} \simeq0$ and $\vq_y^{~2}\simeq0$,
whence $\V{\vq_x,\vq_x,\vq_y}$ remains the only non-zero Yukawa
coupling.\ping

Consider now the $(a,c)$-sector of the \LGO\ $\cal Q$. For a product
of three charge-$({-}1,1)$ states, the Yukawa coupling is
\eqn\eXXX{ \k_{ijk} ~~=~~ \bra{0,0}
            \ket{{+}3,0}^7 \ket{0,{-}3}^7 \ket{{-}1,1}^{\ell_i}_i
             \ket{{-}1,1}^{\ell_j}_j \ket{{-}1,1}^{\ell_k}_k~, }
While the $J_0,\Jb_0$-charges are automatically balanced, the
twist-number selection rule in the $(a,c)$-sector requires
\eqn\eTac{ \ell_i+\ell_j+\ell_k ~\id~ 2\pmod{d}~, }
with $d=8$ here. Therefore, the only non-zero Yukawa coupling is
\eqn\eQY{ \bra{0,0} \ket{{+}3,0}^7 \ket{0,{-}3}^7 \ket{{-}1,1}^3
           \ket{{-}1,1}^3 \ket{{-}1,1}^4~. }
This implies that
\eqn\eXXX{ \ket{{-}1,1}^3 ~\buildrel{M}\over\sim~ \vq_x~,\qquad
           \ket{{-}1,1}^4 ~\buildrel{M}\over\sim~ \vq_y }
are two mirror-pairs.

Next, comparing with the geometrical results in~\eGGG{a,b}, we see
that $\ket{{-}1,1}^3 \sim J_x$ while $\ket{{-}1,1}^4 \sim J_y$ is the
only possible identification. The fact that $\V{J_x,J_x,J_x}\ne0$ in
the geometrical calculation shows that the world-sheet instanton
corrections indeed are not negligible: they in fact cancel out the
classical (tree-level, topological) contribution to $\V{J_x,J_x,J_x}$.

The fact that the instanton effects conspire so as to cancel the
`topological' contribution to some of the Yukawa couplings is easily
seen to be the consequence of a quantum symmetry, that is, a special
(and possibly singular) choice of the K\"ahler class for which the
symmetry used for the GSO-type projection is in fact an isometry.
Such cancellations were to be expected, much as symmetries of the
superpotential relate Yukawa couplings in the $(c,c)$-sector and
often cause some of them to vanish.

Note that there does exist a {\it singular} limit in which the
geometrical Yukawa coupling $\V{J_x,J_x,J_x}$ vanishes but
$\V{J_x,J_x,J_y}$ remains non-zero. We observe that $\V{J_x,J_x,J_x}$
may be identified with the real 6-volume of the projection of the \CY\
3-fold $\cal Q$ on $\IP^4$, while the projection of $\V{J_x,J_x,J_y}$
on $\IP^4$ has the interpretation of a 4-volume.
Thus, for $\V{J_x,J_x,J_x}$ but not $\V{J_x,J_x,J_y}$ to vanish, it
would suffice to collapse $\IP^4$ in one (complex) dimension only.
It seems possible to think of the world-sheet instanton contributions
as {\it effectively} distorting the originally homogeneous geometry of
$\IP^4$ in such an amusing fashion.

Finally, suffice it here to state that the Koszul representatives for
$J_x$ and $J_y$ are simply two scalars, consistent with the fact that
$J_x$ and $J_y$ are invariant under any holomorphic symmetry of $\cal
Q$. (To prove this invariance, simply use the Fubini-Study K\"ahler
forms: $\vd\db\log\|X\|^2$ and $\vd\db\log\|Y\|^2$.) Thus, the Koszul
computation by itself does not provide enough information about the
pull-backs of the K\"ahler forms of the embedding $\IP^n$'s. We
therefore rely on the geometrical calculation of Yukawa couplings and
take an appropriate limit to recover the \LGO\ result; we will call
this the `\LGO\ limit of the Koszul (geometrical) calculation'.

This degree of freedom simply means that, unlike the \LGO\
computation, the Koszul computation does not refer to a particular
but rather to a general choice of the K\"ahler class on the
manifold~\refs{\rBeast, \rMatter}. Consequently, agreement with the
geometrical (large size) or \LGO\ (`small' size) results is obtained
only upon fixing by hand this degree of freedom of the Koszul
computation.

Alternatively, the K\"ahler class variation which takes us away from
the limit in which $\V{J_x,J_x,J_x}=0$ may easily be studied in the
\LGO\ framework, as a deformation of the mirror model~\eQPM. For
example, a multiple of a general quartic polynomial in
$\hat{X}_3,\hat{X}_4,\hat{X}_5$ may be added to $P_{\ha{\cal
Q}}(\hat{X},\hat{Y})$. Thereupon, the result $\vq_y{}^2 \simeq 0$
will still hold, but $\vq_x{}^3$ will no longer vanish in general. Of
course, other deformations will result in a more general change and
render all Yukawa couplings nonzero. In this way we can interpolate
between the \LGO\ results, where some of the characteristic sizes are
small, and the geometric results where all characteristic sizes are
sufficiently big for the world-sheet instanton effects to be
negligible.

\subsec{Another model}\noindent
The first sample model had a somewhat atypical $(a,c)$-sector, in
that the only field representatives were actually two $(a,c)$-vacua.
The next model will have a richer $(a,c)$-sector and in fact will
typify the general case.

Consider the family of \CY\ 3-folds one member of which was used in
Ref.~\rRolf\ to construct a 3-generation manifold:
\eqn\eMDf{
{\cal M} \in \K{3\cr2\cr}{3&1\cr0&3\cr}^{8,35}_{-54}~~:
 ~~\cases{f(x)   = f_{abc}\, x_a\, x_b\, x_c &= 0~,\cr
          g(x,y) = g_{a\,\a\b\g}\, x_a\, y_\a\, y_\b\, y_\g &= 0~.\cr}
}
The most general variations of $f(x)$ and $g(x,y)$ are spanned by 60
monomials. Linear reparametrizations on $\CP3 \times \CP2$ are
elements of $PGL(4;\IC) \times PGL(3;\IC)$ which has $15+8=23$
generators. Taking into account two overall rescalings of the two
homogeneous equations~\eMDf, the 60 monomials form $60-25=35$
independent classes; in fact, all of $H^1({\cal T})$ is represented by
deformations of the defining polynomials $f(x)$ and $g(x,y)$~\rPDM.
So, the $(c,c)$-sector here is simpler than in the first example in
that there are only untwisted charge-$(1,1)$ states. Of course, the
Chiral and the Jacobian ring structures agree, as was well known.\ping

For generic choices of $f(x)$ and $g(x,y)$, the associated \LGO\ is
again well defined simply by setting the superpotential to be
$P_{\cal M}=f(X) + g(X,Y)$~\rBVW. We again easily find:
\eqn\eQCh{ {\cal M} ~:~
 \q_X=q_X=\frc13~, \quad \q_Y=q_Y=\frc29~, \qquad \Q = \ZZ_9~. }
The charges of all 9 Ramond ground states are again found
using~\eRCh; spectral flow then provides the Neveu-Schwarz vacua.
For completeness, Table~4 lists all five classes of vacua.

Again, there exists a special choice of such a superpotential,
\eqn\eMP{ P_{\cal M} =
 \sum_{\a=1}^3\big( X_\a{}^3 + X_\a\, Y_\a{}^3 \big) + X_0{}^3~, }
and the limit of the renormalization flow of this \LGO\ results in a
tensor product $(E_7)^3(A_2)$ of minimal models~\rGepnerIII. This
choice corresponds to setting
\eqn\eGCh{ f_{abc}       \define \cases{1 & $a=b=c$,      \cr
                                        0 & otherwise,    \cr}
   \qquad  g_{a\,\a\b\g} \define \cases{1 & $a=\a=\b=\g$, \cr
                                        0 & otherwise,    \cr}}
in Eq.~\eMDf.

\subsec{Field representatives}\noindent
For $H^1({\cal M},\CM)$, the Koszul calculation produces
\eqn\eMJs{ H^1({\cal T}\con)~ \sim ~\{J_x\} \oplus \{J_y\}
 \oplus \{ \e^{abcd}\vq^{(ij)} ~:~ f_{ijk}\vq^{(ij)}=0 \}_6~. }
This time, besides the pull-backs of the K\"ahler forms $J_x$ and
$J_y$ of $\IP^3$ and $\IP^2$, represented by two scalars, we also
obtain 6 tensors where the totally antisymmetric symbol $\e^{abcd}$
is readily factored out. Note that the scaling charge of the
holomorphic volume form $(x\rd^3x)(y\rd^2y)$ is 2. The scaling charge
of the two scalars, $J_x$ and $J_y$, is of course zero, while the
scaling charge of $\e^{abcd}\vq^{(ij)}$ is $6\inv3=2$, which again is
zero modulo the scaling charge of the holomorphic volume form. This
was expected, as all these field representatives correspond to twisted
states in the \LGO\ framework.

The dual cohomology, $H^2({\cal T})$, is then
spanned by two copies of the holomorphic volume element on $\IP^3
\times \IP^2$:
\eqn\eXXX{ (x\rd^3x)(y\rd^2y)~ \define
               ~(x_0\rd x_1\rd x_2\rd x_3) (y_1\rd y_2\rd y_3)~, }
and six independent tensor coefficients in
\eqn\eMsD{ \{ \e^{\a\b\g}\vq_{(ij)}~ \simeq \e^{\a\b\g}\vq_{(ij)}~
 + \e^{\a\b\g} \l^k f_{ijk}\}_6~. }
Note again that the scaling charges of all these eight field
representatives are zero, modulo the scaling charge of the total
holomorphic volume form.

The latter tensor coefficients, $\vq_{(ij)}$, are themselves naturally
dual to monomials $X_i X_j$, which transform like the original tensor
coefficients $\vq^{(ij)}$. Thus we have
\eqn\eXXX{ \vq^{(ij)} ~\buildrel*\over\sim~ \vq_{(ij)}
                      ~\buildrel*\over\sim~ X_i X_j~,
  \qquad\Longrightarrow\qquad
           \vq^{(ij)} \approx X_i X_j~. }
So, again, we are looking for a ``polynomial'' representative for
$\e^{abcd}$, that is, one which is dual to $\e^{\a\b\g}$.\ping

As remarked above and which is now  obvious from Table~4, the only
integral charge-$(1,1)$ states come from the untwisted
$(c,c)$-sector. In the $(a,c)$-sector, however there are two
charge-$({-}1,1)$ vacua, with $\ell=3,5$ and there is also the vacuum
$\ket{{-}\frc53,\inv3}^4$, which becomes a charge-$({-}1,1)$ state
when multiplied with quadrics in $X_a$. Indeed, $\ell=4$ in the
$(a,c)$-sector stems from $\ell=3$ in the Ramond sector and there the
$X_a$ but not the $Y_\a$ are left invariant by the action of
$\Q=\ZZ_9$. Reducing furthermore modulo the ideal generated by the
gradients of the superpotentials, this yields six states of the form
\eqn\eXXX{ \vq(X)\ket{{-}\frc53,{+}\inv3}^4~,\qquad
           \vq(X) \cong \vq(X) + \l^a\vd_a\, f(X)~. }
It is immediate that such quadrics are dual to the rank-2 tensor
$\vq_{(ij)}$ in~\eMsD, that is, the quadrics $\vq(X)$ here may be
identified with the tensors $\vq^{(ij)}$ in~\eMJs. Again, twisted
vacuum $\ket{{-}\frc53,{+}\inv3}^4$ is tentatively identified with
the $\e^{abcd}$ symbol and we seek a ``polynomial'' representative
for it.\ping

By switching to $\ha{\cal M}$, the mirror model of $\cal M$, the
eight charge-$({-}1,1)$ states will all acquire charge-$(1,1)$
counterparts. For the special choice of the superpotential
$P_{\cal M}$ as in~\eMP, the superpotential is self-transposed and
mirror is obtained simply as a quotient of $\cal M$ itself~\rGCM.
As this is not the case in general, we shall place a `hat' on the
coordinates when they refer to $\ha{\cal M}$.
It is easy to check that (the pull-backs of) the K\"ahler forms $J_x$
and $J_y$ have the following mirrors:
\eqn\eXXX{ \vq_x \define \hat{X_0} \hat{Y_1} \hat{Y_2} \hat{Y_3}~,
  \qquad   \vq_y \define \hat{X_1} \hat{X_2} \hat{X_3}~, }
where again the choice of the labeling involved a little forethought.
In addition, the mirrors of $X_a X_b\ket{{-}\frc53,{+}\inv3}^{4}$ will
be twisted $(c,c)$-states, however, only consisting of  twisted
$(c,c)$-vacua\ft{We thank M.~Kreuzer for pointing out an error in an
early version of the paper.}.

Note that the above two representatives, $\vq_x$ and $\vq_y$, are the
two factors of the `fundamental monomial' $\prod_a\hat{X}_a
\prod_\a\hat{Y}_\a$ (see Ref.~\rSL) which are invariant under the
$\ZZ_3\times\ZZ_9$ symmetry---the action of which we must divide out
to obtain the mirror $\ha{{\cal M}}$. Also, $\vq_x$ and $\vq_y$ are
nontrivial deformations of $\cal M$ (upon un-hatting the coordinate
fields) as well as of the mirror model (with the hats on), and will
remain so also away from the specially symmetric choice~\eMP~\rSL.

\subsec{Yukawa couplings}\noindent
The twist-number selection rule~\eTac\ alone dictates that
\eqna\eYac
$$ \eqalignno{
 &\kern-\parindent
\V{J_x,J_y,J_y}~ \sim
 ~\bra{0,0} \ket{{+}3,0}^7 \ket{0,{-}3}^7 \ket{{-}1,1}^3
             \ket{{-}1,1}^5 \ket{{-}1,1}^5\,, \qquad  &\eYac{a} \cr
\noalign{\vglue1mm}
 &\kern-\parindent
   \V{\e^{....}\vq^{(ij)},\e^{....}\vq^{(pq)},J_y}~ \sim\hfill   \cr
\noalign{\vglue-1mm}
 &\ifx\answ\bigans\kern-1.0\parindent\else\kern-2.0\parindent\fi
  \bra{0,0} \ket{{+}3,0}^7 \ket{0,{-}3}^7
  (X_i X_j\ket{{-}\frc53,\inv3}^4) (X_p X_q\ket{{-}\frc53,\inv3}^4)
           \ket{{-}1,1}^5 \qquad                 &\eYac{b} \cr}
$$
are the only non-vanishing $(a,c)$-sector Yukawa couplings in the
\LGO\ limit.

As in the previous case, away from the \LGO\ limit, the geometrical
computation also yields $\V{J_x,J_x,J_y}\ne0$. Again, the
$\V{J_x,J_x,J_y}\ne0$ result can be obtained also in the \LGO\
framework, by calculating $\V{\vq_x,\vq_x,\vq_y}\ne0$ for a deformed
model of $\ha{\cal M}$, e.g., one in which a multiple of $\vq_x$
has been added to $P_{\ha{\cal M}}$. This happens in exact
parallel to the previous example and we obtain that
\eqn\eXXX{ \ket{{-}1,1}^3 ~\buildrel{M}\over\sim~ \vq_x~,\qquad
           \ket{{-}1,1}^5 ~\buildrel{M}\over\sim~ \vq_y }
are mirror-duals. In fact, since both $X_0Y_1Y_2Y_3$ and $X_1X_2X_3$ are
invariant under the $\ZZ_3\times\ZZ_9$ quotient which gives the mirror,
they are present both in the original and the mirror model, so that
$\ket{{-}1,1}^3$ and $\ket{{-}1,1}^5$  may be identified with
$X_0Y_1Y_2Y_3$ and $X_1X_2X_3$ respectively as their  monomial
representatives. The `top element'
$Q_{\cal M} = 2{\cdot}3^7 7^3\,X_0 X_1^2Y_1 X_2^2Y_2 X_3^2Y_3$ is
inherited by the mirror model, since $\ha{\cal M} = {\cal
M}/{\mit\D}$ and only those quotients are allowed where $\mit\D$
leaves $Q_{\cal M}$ invariant. Clearly,
$\vq_x \vq_y^{~2} \propto Q_{\ha{\cal M}} = 2{\cdot}3^7
7^3\,\hat{X}_0 \hat{X}_1^2\hat{Y}_1 \hat{X}_2^2\hat{Y}_2
\hat{X}_3^2\hat{Y}_3$, which in fact equals $Q_{\cal M}$.

The type of the Yukawa coupling~\eYac{b} did not occur in the previous
model and provides new information here. From \eYac b\ and the discussion
in the above paragraph we find that
\eqn\eXXX{ \big(\ket{{-}\frc53,\inv3}^4\big)^2
                           ~ = ~Y_1 Y_2 Y_3 \big(\ket{0,0}^0\big)^2 }
must hold. In other words, $\sqrt{Y_1 Y_2 Y_3}$ may be regarded
as the `polynomial' representative of the twist-field producing
$\ket{{-}\frc53,\inv3}^4$ from $\ket{0,0}^0$
\ft{The analysis can also be carried out by considering the
$(c,c)$-ring of the mirror $\ha{\cal M}$ only. By using the selection rules
and the ringstructure, in particular the existence of the top element
$Q_{\ha{\cal M}}$, one finds that the six twisted (`non-polynomial')
$(1,1)$ states are given by
$\hat X_a\hat X_b\sqrt{\hat Y_1 \hat Y_2 \hat Y_3}_{~a\ne b}$.}.

Turning back to the Koszul computation, the analogue of this result is
not hard to prove. That is, the factor $\e^{abcd}$ in the field
representative~\eMJs\ may be assigned the `polynomial' representative
$\sqrt{Y_1 Y_2 Y_3}$. To see this, consider the square of the
$\e^{\a\b\g}$ from the six dual representatives in~\eMsD.
Through contraction with the defining tensors $g_{a\,\a\m\l}
g_{b\,\b\n\k} g_{c\,\g\r\s}$, the square of the antisymmetric symbol,
$\e^{a\b\g} \e^{\m\n\r}$ is dual to
\eqn\eDVQ{ X_a Y_\l X_b Y_\k X_c Y_\s
  ~~\tooo{{\rm~~Eq.~\eGCh~~}}~~\check{\vq}_y{\cdot}Y_1 Y_2 Y_3~, }
where the result on the r.h.s.\ of the arrow in~\eDVQ\ has been
obtained with the simple choice of the defining tensors~\eGCh, and
$\check{\vq}_y=X_1X_2X_3$ is the un-hatted version of $\vq_y$.

Recall finally that in the Koszul computation, the pull-backs of the
K\"ahler forms enter as scalars, rather than tensors dual to
$\check{\vq}_x=X_0Y_1Y_2Y_3$ or $\check{\vq}_y=X_1X_2X_3$. Therefore,
the overall factor of $\check\vq_y$ in Eq.~\eDVQ\ was necessary just
so as to make the Yukawa coupling~\eYac{b} non-zero in the Koszul
framework.\ping

It is then clear that the \LGO\ and the $\V{J_x,J_x,J_y}\to0$ limit
of the Koszul calculation (as discussed in the previous example)
predict the same Yukawa couplings to vanish. We may now address the
issue of the ratio between all the non-zero Yukawa couplings~\eYac{a}
and~\eYac{b}.  In the \LGO\ framework, this is a definite number,
equal to 1 with the current definitions of the field representatives.
In the Koszul computation, the coupling~\eYac{a} is simply a
product of scalars and is therefore already a scalar quantity, which
can be set equal to the value of~\eYac{b}.

The ratio of the Yukawa couplings is therefore in both
frameworks a matter of normalization of the field representatives
and can clearly be made to agree. Of course, in doing so, we had to
take the \LGO\ limit of the Koszul results, whence $\V{J_x,J_x,J_y}$
was made to vanish as discussed in detail in the previous example.

\subsec{The general case}\noindent
As for the $(c,c)$ ring, we would like to extend the results for the
$(a,c)$ ring to the situation when there is no (known) \LGO\
corresponding to the complete intersection manifold description.

Let ${\cal M}$ be defined as in section~{\it3.5}.  The tensor
representatives for the elements in $H^1({\cal M},\CM)$ are  of the
form $\e^{\ldots} \f^{\ldots}$. Except for some exceptional
cases~\refs{\rBeast, \rPDM}, here are also $N$ scalars $J_i$ each one
being the pull-back of the corresponding K\"ahler form on $\CP{n_i}$.
In order to  derive the monomial representatives for the various
$\e^{\ldots}$'s, we consider the dual cohomology $H^2({\cal T})$. The
dual of the  $\e^{\ldots} \f^{\ldots}$ are given by $\Tw\e^{\ldots}
\j_{\ldots}$, where $\e^{\cdots}$ and $\Tw\e^{\cdots}$ are dual in the
sense that their product is the total $\e$-symbol for the whole field
space. As for the $(c,c)$ ring then, we find the dual monomial
representatives of the $\Tw\e^{\ldots}$ by considering the square,
$\Tw\e^2$, and contract with the appropriate $f^{(m)}_{\ldots}$ and
$x^{(i)}_j$. The square root of the so entered product of
$x^{(i)}_j$'s is the `monomial' representative for $\e^{\ldots}$.

We also need to identify the $J_i$ with a certain `canonical' product
of coordinates. To that effect we have to consider the `canonical'
deformations of the mirror manifold ${\cal W}$. However, only in a
very few cases  do we know what the mirror is.  As for the class of
weighted projective spaces one may guess that ${\cal W}$ will be a
quotient of some other complete intersection $\Tw{\cal M}$. For a
large class of models of the former type, it has been argued that the
mirror model is obtained as a quotient of a model based on the
transpose of the defining polynomial~\rGCM. It is tempting to say that
the same will be true for the whole class of complete intersection
\CY\ manifolds~\rCYCI. Thus the mirror manifold would be ${\cal
M}^T/H$ where ${\cal M}^T$ is the transposed complete intersection and
$H$ is the quotient group necessary to produce ${\cal W}$. By ${\cal
M}^T$ we mean to take the transpose of the degree matrix, given by the
defining equations. In order for the transposed theory to be a \CY\
manifold, one will have to consider the hypersurface to be embedded in
a different product of weighted projective spaces such that the
condition of vanishing first Chern class is satisfied in each of the
projective spaces. We hope to return shortly with a proof of the above
mirror hypothesis.

{}From each defining equation (there are now $N$ of them) we have a
`fundamental deformation' much in the same way as in the case of just
one (weighted) projective space~\rSL. These $N$ monomials we identify
as the $J_i$. At this point we have a complete set of coordinate
representatives for the $H^1({\cal M},\CM)$. The computation of the
Yukawa couplings follows in the same way as in the $(c,c)$ ring.

\newsec{Concluding Remarks}\noindent
To complete proving the equivalence of the various computations of
the Yukawa couplings, up to the unspecified K\"ahler class in the
Koszul computation, one has to address the issue of field
representative normalizations, i.e., the kinetic terms.

Besides studying this problem by explicit computation, model by model,
we are aware of two general approaches. On one hand, a powerful
machinery is currently being developed for the calculation of
so-called periods $\int_\g\W$ of the holomorphic 3-form\ft{We thank
P.~Candelas and collaborators for sharing their results prior to
publication.}. These quantities turn out to completely determine both
the `un-normalized' Yukawa couplings and also the kinetic terms for the
effective {\bf27} and {\bf27\con} fields. This study will be reported
elsewhere.

Another method is based on the analysis of Ref.~\rCeGiPa, where the it
was proven---for Fermat type polynomials, i.e., tensor products of
$A_k$-type minimal models---that the \LGO\ analysis can be extended to
provide the kinetic terms identical to those of the exact
superconformal models. The key point is that the Witten vacuum
wave functions can be completely determined by solving certain
`Schr\"odinger equations', and that the usual normalization of these
wave-functions precisely reproduces the kinetic terms dictated by the
exact super\CFT. Of course, the general case involves more than just
the $A_k$-type minimal models. Classifications of admissible
superpotentials for \LGO{s}~\refs{\rGCM, \rMax} enable us to isolate,
in addition to the Fermat type, two more classes of polynomials and
the corresponding Wess-Zumino models and their analysis will be
reported shortly.

In summary, we have extended the results of Ref.~\rCQG, and
demonstrated a complete agreement between the Koszul computation and
the \LGO\ computation of the Yukawa couplings. This agreement extends
over both twisted and untwisted and both $(c,c)$ and $(a,c)$-sectors.
To demonstrate this, we have used `ineffective splitting' and the
mirror map. Next, the `twist fields' which create twisted vacua
from untwisted ones have successfully been identified with the
$\e^{\cdots}$ symbols of the Koszul computation. Moreover, these have
been assigned representatives which are square roots of polynomials
and which allow a straightforward generalization of the well-known
Yukawa coupling formula~\rPhilip---for all {\bf27}'s and {\bf27\con}'s.
In fact, we find (modulo our mirror conjecture for the complete
intersection \CY\ manifolds) that the Koszul computation can be
translated into the existence of a polynomial $(c/a,c)$ ring even
when the underlying $N=2$ superconformal field theory is not known.

Finally, we wish to emphasize that neither the Koszul computation, nor
the \LGO\ techniques depend on criticality (the \CY\ condition),
although the relation between the two will become more involved
off-criticality. Also, these methods may be extended to field theories
in higher dimensions, since they pertain to the analysis of
supersymmetric vacuum wave-functions which exhibit a high degree of
universality~\rCeGiPa.

 %
\vfill
{\bf Acknowledgments}:
One of us (P.B)  would like to thank the Theory Division at CERN for
hospitality where this work was initiated. P.B. acknowledges the
support of the America-Scandinavian Foundation, the Foundation
Blanceflor-Boncompagni-Ludovisi n\'ee Bildt, the Fulbright Program, a
University of Texas Fellowship, the NSF grants  PHY 8904035 and PHY
9009850 and the Robert~A.~Welch foundation.
T.H. was supported in part by the DOE grant DE-FG02-88ER-25065 and
would also like to thank the Department of Mathematics of the National
Tsing-Hua University at Hsinchu, Taiwan, for the warm hospitality
during the time when part of this research was completed.

\vfill\eject

 %
\appendix{A}{Gauge-Theoretic Introduction to Koszul Computation}
 \noindent
The Koszul computation may be formulated as an extension of the
familiar idea of gauge transformation and gauge invariance. In fact,
some treatments of (second class) gauge constraints in the physics
literature are remarkably similar to the standard treatment of the
Koszul complex in the mathematical literature. The modest basic facts
are reviewed here and we refer to Refs.~\refs{\rBeast, \rMichael,
\rMatter} for more details and a more complete account.

The Koszul computation in general refers to the task of computing
certain cohomology groups of a manifold $\cal M$, which is given as a
submanifold in a better known $\cal X$ as the vanishing set of a
system of constraints. To this end, consider  the simple case in which
$\cal M$ is  a hypersurface in $\cal X$, given globally as the
zero-set of a single constraint equation $f(x)=0$. It is rather
standard that local functions $\f(x)$ on $\cal M$ are obtained as a
restriction to $x\in\cal M$ of local functions $\F(x)$ on $\cal Y$. In
doing so, however, $\f(x)$ will be well defined only up to appropriate
$\l(x)$-multiples of $f(x)$:
\eqn\eXXX{ \f(x)~~ \cong ~~\F(x)~ + ~\l(x){\cdot}f(x)~,\qquad
           x\in{\cal M}~, }
simply because on $\cal M$, $f(x)$ vanishes and so do the
$\l(x)$-multiples of $f(x)$. Of course, the functions $\l(x)$ are
chosen so that the product $\l(x){\cdot}f(x)$ would transform just as
$\f(x)$ does. In other words, we consider the gauge-transformation
\eqn\eXXX{ \d \F(x)~~ = ~~\l(x) {\cdot} f(x) }
in the space of functions on $\cal X$. All such gauge-transforms of a
function will agree on ${\cal M} \subset \cal X$ and form the
equivalence class which represents a single function on $\cal M$.

The same principle applies equally well to tensors, provided we take
into account the fact that a component of a vector is (anyway) well
defined only up to linear reparametrizations. Since a tangent vector
of $\cal M$ is also a tangent vector of $\cal X$, it is clear that the
former is undefined up to multiples of gradients of the defining
function.

Finally, when looking for various harmonic forms on $\cal M$, it must
be realized that the {\it differential} conditions of closedness and
exactness,  $\rd\w=0$ and $\w=\rd\a$ respectively, have to be
appropriately restricted from $\cal X$ to $\cal M$.

The Koszul computation offers a purely algebraic analogue to all
these requirements; one never has to solve any differential equation.
In a sense, the Koszul computation is like an integral transform,
which turns a system of differential and algebraic equations into a
purely algebraic system.

There are three basic stages of the Koszul computation.
\item{1.} A tensor-valued form on $\cal M$ has to be specified in
          terms of tensor-valued forms on $\cal X$, restricted however
          as a function to ${\cal M} \subset \cal X$;
\item{2.} The restrictions to $\cal M$ of tensor-valued forms on
          $\cal X$ have to be specified in terms of unrestricted
          forms on $\cal X$;
\item{3.} Finally, all required forms on $\cal X$ have to be specified
          in a fashion which will be convenient for calculations.

For this last stage, the Bott-Borel-Weil theorem guarantees that all
required forms can be given in terms of $U(N)$-representations if the
embedding space is chosen to be a product of flag-spaces of the form
$U(N)/\prod_i U(n_i)$, $\sum_i n_i=N$.

For the first stage, one uses the well-known fact that the gradient of
$f(x)$ provides a normal to the hypersurface ${\cal M}\subset\cal X$.
Thus, at any particular point of ${\cal M}\subset\cal X$, a tangent
vector of $\cal X$ decomposes into a vector tangent to $\cal M$ and a
multiple of $\nabla f(x)$. This relation is then iterated for tensors
of higher rank and/or when $\cal M$ is an intersection of two or
more hypersurfaces, that is, the simultaneous zero-set of a system of
constraints rather than a single one.

In the second stage, the components of the various tensors are
treated as functions and are taken modulo appropriate multiples of
(gradients of) the defining function. If the subspace $\cal M$ is
defined as the simultaneous zero-set of more than one constraint,
this passage to gauge-equivalent classes can be done taking one
constraint at a time. However, there is a subtlety in doing so and
this is the basic reason for the growing complexity of the Koszul
computation. Suffice it here to consider a case with two constraints,
$f(x)=0$ and $g(x)=0$.

Since both $f(x)$ and $g(x)$ are set to zero on
${\cal M}\subset\cal X$, a function on $\cal M$ is given as a class
of function on $\cal X$, defined up to the gauge-transformation
\eqn\eGTr{ \d \F(x)~~ =
          ~~\l_f(x){\cdot}f(x) + \l_g(x){\cdot}g(x)~,
          \qquad x\in{\cal M}~, }
where $\l_f(x)$ and $\l_g(x)$ are suitably transforming functions.
It would thus appear that these two functions are arbitrary and
independent and that the space of functions is a quotient
\eqn\eQUO{ \big\{\, \f(x) \,\big\}~~ =
  ~~\Big\{\, \big\{ F(x) \big\} \,\Big/\,
              \big\{ \l_f(x){\cdot}f(x) \oplus
                      \l_g(x){\cdot}g(x) \big\} \,\Big\}~. }
Notice, however, that by choosing
\eqn\eXXX{ \l_f(x)~ = ~{-}\m(x)g(x)~,\qquad
           \l_g(x)~ = ~{+}\m(x)f(x)~, }
the gauge-transformation of $\F(x)$ vanishes trivially, {\it even away
from} $\cal M$. Therefore, too much has been divided out in the
quotient~\eQUO; the equivalence relation in~\eGTr\ is too big. The
gauge transformation~\eGTr\ is itself subject to a second-level
gauge-transformation
\eqn\eGTR{ \d\l_f(x)~ = ~{-}\m(x)g(x)~, \qquad
           \d\l_g(x)~ = ~{+}\m(x)f(x)~. }

In a field theory where the constraints $f(x)=0$ and $g(x)=0$ are
implemented by means of Lagrange multipliers, the second-level
gauge-transformation~\eGTR\ would act on the Lagrange multipliers of
$f(x)$ and $g(x)$. Just as in a BRST treatment there would be
introduced ghost fields of $\l_f(x)$ and $\l_g(x)$, there would now
also be introduced ghost-for-ghost of $\m(x)$.

Clearly, with more than two constraints, there will be gauge
invariances of higher levels, the total level number equaling the
total number of independent constraints. In the Koszul computation,
these hierarchies of gauge-transformations are systematically and
effectively taken care by the so-called spectral sequence, and we
refer the interested Reader to Refs.~\refs{\rBeast, \rMichael,
\rMatter} for details of this technique.

\appendix{B}{On Ideals}\noindent
It was shown in Ref.~\rCQG\ that the Koszul ideal and the ideal of the
Chiral ring structure are in fact equal. For the sake of completeness
and also to illustrate several simplifications which take place when a
\LGO\ formulation is possible, we compute this for the particular
example studied in section~2.

For simplicity let us only consider the ideal that occurs in the
 Chiral ring. The superpotential is defined to be the sum
\eqn\eXXX{ P_{\cal Q}~ \define ~f(X) + g(X,Y)~. }
Clearly, for this to make sense, one has to be able to assign
scaling charges to the chiral fields $X_a$ and $Y_\a$ such that
$P_{\cal Q}$ scales quasihomogeneously; in general, this may be an
important obstruction to construct \LGO{s} corresponding to a given
\CY\ manifold~\refs{\rElusive, \rCQG}.

Thereupon, the ideal in the Chiral ring is generated by the gradients
of $P_{\cal Q}$:
\eqn\eLGI{ \Im[\vd P_{\cal Q}] ~~\sim~~
            \big\{\, (\vd_a f+\vd_a g) \,,\, (\vd_\a g) \,\big\}~, }
where, of course, $\vd_a\define{\vd\over\vd X_a}$ and
$\vd_\a\define{\vd\over\vd Y_\a}$.

Since the forms we are interested in are all (co)tangent-valued, they
will be well defined up to addition of multiples of gradients of the
defining functions $f(x)$ and $g(x,y)$ in~\eQDf. In fact, since these
functions are homogeneous, the gauge-transformation of the type~\eGTr\
will also be generated by addition of multiples of the gradients of
$f(x)$ and $g(x,y)$.
{Na\"\ii}vely, therefore, one would take that the various
tangent-valued forms are defined up to additive elements of the ideal
\eqn\eNI{ \Im[\vd f, \vd g] ~~\sim~~ \big\{\, (\vd_a f)
            \,,\, (\vd_a g) \,,\, (\vd_\a g)  \,\big\}~, }
consisting of all multiples of the indicated gradients. Comparing the
ideal~\eNI\ with~\eLGI, it is quite obvious that this {na\"\ii}ve
ideal is too big. In fact, we have the same problem as with the
quotient~\eQUO. The complete Koszul computation takes care of the
second-level gauge-transformation~\eGTR, which has not been taken
into account in~\eNI.

In general, the discrepancy is rather involved. However, in cases
when the \LGO\ formulation is possible, that is, when the fields
$X_a,Y_\a$ can be assigned scaling charges such that $f(x)$ and
$g(x,y)$ have the {\it same} homogeneity, the Koszul ideal is simply
the na\"\ii ve one in~\eNI, taken itself modulo multiples of the
generator of the second-level gauge-equivalence $\vd_a(f-g)$.
In the present case, $\vd_\a f=0$ trivially, so that $\vd_\a(f-g)$
does not generate a redundant relation in the Koszul ideal. Since
\eqn\eXXX{ \bigg\{ {\l^a_f{\cdot}\vd_a f \oplus \l^a_g{\cdot}\vd_a g
                      \over \m{\cdot}[\vd_af - \vd_ag]} \bigg\}
 ~~\approx~~ \big\{\, \n{\cdot}[\vd_af + \vd_ag] \,\big\}~, }
we have proven that the Koszul ideal equals the Landau-Ginzburg one,
occurring in the Chiral ring of the \LGO.

With more than two defining functions for a \CY\ manifold---if a
\LGO\ can be defined---the same equality will hold, although proving
it will be rather more involved technically. A simple example with
three defining functions is presented in Ref.~\rCQG.
\vfill
\eject

 %
\def\1{\vphantom{\frc12}}
\vglue0mm
\vfill
\vbox{
$$\vbox{\offinterlineskip
\hrule height1pt
\halign{
&\vrule width1pt#&\strut~~#\hfil~~&\vrule#
                  &\strut~~#\hfil~~&\vrule#
                   &\strut~~#\hfil~~&\vrule#
                    &\strut~~#\hfil~~&\vrule#
                     &\strut~~#\hfil~~&\vrule width1pt#\cr
height2pt&\omit&&\omit&&\omit&&\omit&&\omit&\cr
&\hfil$(a,a)$&&\hfil$(c,c)$&&\hfil Ramond &
&\hfil$(a,c)$&&\hfil$(c,a)$&\cr
height2pt&\omit&&\omit&&\omit&&\omit&&\omit&\cr
\noalign{\hrule\vskip1pt\hrule}
height3pt&\omit&&\omit&&\omit&&\omit&&\omit&\cr
&$\ket{\1{-}3,{-}3}^0$&
&$\2{\2{\ket{\10,0}^0}}$&
&$\ket{\1{-}\frc32,{-}\frc32}^0$&
&$\ket{\1{-}3,0}^1$&
&$\ket{\10,{-}3}^7$&\cr
height3pt&\omit&&\omit&&\omit&&\omit&&\omit&\cr
\noalign{\hrule}
height3pt&\omit&&\omit&&\omit&&\omit&&\omit&\cr
&$\ket{\1{-}3,0}^1$&
&$\ket{\10,{+}3}^1$&
&$\ket{\1{-}\frc32,{+}\frc32}^1$&
&$\ket{\1{-}3,{+}3}^2$&
&$\ket{\10,0}^0$&\cr
height3pt&\omit&&\omit&&\omit&&\omit&&\omit&\cr
\noalign{\hrule}
height3pt&\omit&&\omit&&\omit&&\omit&&\omit&\cr
&$\ket{\1{-}1,{-}2}^2$&
&$\ket{\1{+}2,{+}1}^2$&
&$\ket{\1{+}\frc12,{-}\frc12}^2$&
&$\2{\2{\ket{\1{-}1,{+}1}^3}}$&
&$\ket{\1{+}2,{-}2}^1$&\cr
height3pt&\omit&&\omit&&\omit&&\omit&&\omit&\cr
\noalign{\hrule}
height3pt&\omit&&\omit&&\omit&&\omit&&\omit&\cr
&$\ket{\1{-}1,{-}2}^3$&
&$\ket{\1{+}2,{+}1}^3$&
&$\ket{\1{+}\frc12,{-}\frc12}^3$&
&$\2{\2{\ket{\1{-}1,{+}1}^4}}$&
&$\ket{\1{+}2,{-}2}^2$&\cr
height3pt&\omit&&\omit&&\omit&&\omit&&\omit&\cr
\noalign{\hrule}
height3pt&\omit&&\omit&&\omit&&\omit&&\omit&\cr
&$\ket{\1{-}\frc{11}4,{-}\frc{11}4}^4$&
&$\2{\2{\ket{\1{+}\frc14,{+}\frc14}^4}}$&
&$\ket{\1{-}\frc54,{-}\frc54}^4$&
&$\ket{\1{-}\frc{11}4,{+}\frc14}^5$&
&$\ket{\1{+}\frc14,{-}\frc{11}4}^3$&\cr
height3pt&\omit&&\omit&&\omit&&\omit&&\omit&\cr
\noalign{\hrule}
height3pt&\omit&&\omit&&\omit&&\omit&&\omit&\cr
&$\ket{\1{-}2,{-}1}^5$&
&$\ket{\1{+}1,{+}2}^5$&
&$\ket{\1{-}\frc12,{+}\frc12}^5$&
&$\ket{\1{-}2,{+}2}^6$&
&$\ket{\1{+}1,{-}1}^4$&\cr
height3pt&\omit&&\omit&&\omit&&\omit&&\omit&\cr
\noalign{\hrule}
height3pt&\omit&&\omit&&\omit&&\omit&&\omit&\cr
&$\ket{\1{-}2,{-}1}^6$&
&$\ket{\1{+}1,{+}2}^6$&
&$\ket{\1{-}\frc12,{+}\frc12}^6$&
&$\ket{\1{-}2,{+}2}^7$&
&$\ket{\1{+}1,{-}1}^5$&\cr
height3pt&\omit&&\omit&&\omit&&\omit&&\omit&\cr
\noalign{\hrule}
height3pt&\omit&&\omit&&\omit&&\omit&&\omit&\cr
&$\ket{\10,{-}3}^7$&
&$\ket{\1{+}3,{+}0}^7$&
&$\ket{\1{+}\frc32,{-}\frc32}^7$&
&$\ket{\10,0}^0$&
&$\ket{\1{+}3,{-}3}^6$&\cr
height3pt&\omit&&\omit&&\omit&&\omit&&\omit&\cr}
\hrule height1pt}$$
\vskip5pt
\noindent
{\bf Table 1}: The Ramond and Neveu-Schwarz vacua of the
               Landau-Ginzburg orbifold $\cal Q$, defined
               in Eq.~\eQDf.}
\ifx\answ\bigansw\vfill
            \else\vfill\goodbreak\vglue0mm\vfill\fi
\vbox{
$$\vbox{\offinterlineskip
\hrule height1pt
\halign{
&\vrule width1pt#&\strut~~#\hfil~~&\vrule#
                  &\strut~~#\hfil~~&\vrule#
                   &\strut~~#\hfil~~&\vrule#
                    &\strut~~#\hfil~~&\vrule#
                     &\strut~~#\hfil~~&\vrule width1pt#\cr
height2pt&\omit&&\omit&&\omit&&\omit&&\omit&\cr
&\hfil$(a,a)$&&\hfil$(c,c)$&&\hfil Ramond &
&\hfil$(a,c)$&&\hfil$(c,a)$&\cr
height2pt&\omit&&\omit&&\omit&&\omit&&\omit&\cr
\noalign{\hrule\vskip1pt\hrule}
height3pt&\omit&&\omit&&\omit&&\omit&&\omit&\cr
&$\ket{\1{-}3,{-}3}^0$&
&$\2{\2{\ket{\10,0}^0}}$&
&$\ket{\1{-}\frc32,{-}\frc32}^0$&
&$\ket{\1{-}3,0}^1$&
&$\ket{\10,{-}3}^7$&\cr
height3pt&\omit&&\omit&&\omit&&\omit&&\omit&\cr
\noalign{\hrule}
height3pt&\omit&&\omit&&\omit&&\omit&&\omit&\cr
&$\ket{\1{-}3,0}^1$&
&$\ket{\10,{+}3}^1$&
&$\ket{\1{-}\frc32,{+}\frc32}^1$&
&$\ket{\1{-}3,{+}3}^2$&
&$\ket{\10,0}^0$&\cr
height3pt&\omit&&\omit&&\omit&&\omit&&\omit&\cr
\noalign{\hrule}
height3pt&\omit&&\omit&&\omit&&\omit&&\omit&\cr
&$\ket{\1{-}2,{-}1}^2$&
&$\ket{\1{+}1,{+}2}^2$&
&$\ket{\1{-}\frc12,{+}\frc12}^2$&
&$\ket{\1{-}2,{+}2}^3$&
&$\ket{\1{+}1,{-}1}^1$&\cr
height3pt&\omit&&\omit&&\omit&&\omit&&\omit&\cr
\noalign{\hrule}
height3pt&\omit&&\omit&&\omit&&\omit&&\omit&\cr
&$\ket{\1{-}1,{-}2}^3$&
&$\ket{\1{+}2,{+}1}^3$&
&$\ket{\1{+}\frc12,{-}\frc12}^3$&
&$\2{\2{\ket{\1{-}1,{+}1}^4}}$&
&$\ket{\1{+}2,{-}2}^2$&\cr
height3pt&\omit&&\omit&&\omit&&\omit&&\omit&\cr
\noalign{\hrule}
height3pt&\omit&&\omit&&\omit&&\omit&&\omit&\cr
&$\ket{\1{-}\frc94,{-}\frc94}^4$&
&$\2{\2{\ket{\1{+}\frc34,{+}\frc34}^4}}$&
&$\ket{\1{-}\frc34,{-}\frc34}^4$&
&$\ket{\1{-}\frc94,{+}\frc34}^5$&
&$\ket{\1{+}\frc34,{-}\frc94}^3$&\cr
height3pt&\omit&&\omit&&\omit&&\omit&&\omit&\cr
\noalign{\hrule}
height3pt&\omit&&\omit&&\omit&&\omit&&\omit&\cr
&$\ket{\1{-}2,{-}1}^5$&
&$\ket{\1{+}1,{+}2}^5$&
&$\ket{\1{-}\frc12,{+}\frc12}^5$&
&$\ket{\1{-}2,{+}2}^6$&
&$\ket{\1{+}1,{-}1}^4$&\cr
height3pt&\omit&&\omit&&\omit&&\omit&&\omit&\cr
\noalign{\hrule}
height3pt&\omit&&\omit&&\omit&&\omit&&\omit&\cr
&$\ket{\1{-}1,{-}2}^6$&
&$\ket{\1{+}2,{+}1}^6$&
&$\ket{\1{+}\frc12,{-}\frc12}^6$&
&$\2{\2{\ket{\1{-}1,{+}1}^7}}$&
&$\ket{\1{+}2,{-}2}^5$&\cr
height3pt&\omit&&\omit&&\omit&&\omit&&\omit&\cr
\noalign{\hrule}
height3pt&\omit&&\omit&&\omit&&\omit&&\omit&\cr
&$\ket{\10,{-}3}^7$&
&$\ket{\1{+}3,{+}0}^7$&
&$\ket{\1{+}\frc32,{-}\frc32}^7$&
&$\ket{\10,0}^0$&
&$\ket{\1{+}3,{-}3}^6$&\cr
height3pt&\omit&&\omit&&\omit&&\omit&&\omit&\cr}
\hrule height1pt}$$
\vskip5pt
\noindent
{\bf Table 2}: The Ramond and Neveu-Schwarz vacua of the
               Landau-Ginzburg orbifold $\Tw{\cal Q}$,
               defined in Eq.~\eQTw.}
\vfill
\eject

\vglue0mm
\vfill
\vbox{
$$\vbox{\offinterlineskip
\hrule height1pt
\halign{
&\vrule width1pt#&\strut~\hfil#&\vrule#
                  &\strut~\hfil#&\vrule#
                   &\strut~\hfil#&\vrule#
                    &\strut~\hfil#&\vrule#
                     &\strut~\hfil#&\vrule width1pt#\cr
&\multispan3\hfil The $\cal Q$ Model \hfil&
 &\vrule width0pt height12pt depth3pt &
 &\multispan3\hfil The $\Tw{\cal Q}$ Model \hfil&\cr
& Koszul \hfil&& Landau-Ginzburg \hfil&& \# \hfil&
& Landau-Ginzburg \hfil&& Koszul \hfil&\cr
height2pt&\omit&&\omit&&\omit&&\omit&&\omit&\cr
\noalign{\hrule\vskip1pt\hrule}
height3pt&\omit&&\omit&&\omit&&\omit&&\omit&\cr
& $\f_{(\a\a\a\b)}$ &
& $X_\a^3 X_\b \ket{\10,0}^0$ &
& 2 &
& $X_\a^3 X_\b \ket{\10,0}^0$ &
& $\f_{(\a\a\a\b)}$ &\cr
height2pt&\omit&&\omit&&\omit&&\omit&&\omit&\cr
\noalign{\hrule}
height3pt&\omit&&\omit&&\omit&&\omit&&\omit&\cr
& $\f_{(\a\a\a\,b)}$ &
& $X_\a^3 X_b \ket{\10,0}^0_{b=3,4,5}$ &
& 6 &
& $X_\a^3 X_b \ket{\10,0}^0_{b=3,4,5}$ &
& $\f_{(\a\a\a\,b)}$ &\cr
height2pt&\omit&&\omit&&\omit&&\omit&&\omit&\cr
\noalign{\hrule}
height3pt&\omit&&\omit&&\omit&&\omit&&\omit&\cr
& $\f_{(aabb)}$ &
& $X_a^2 X_b^2 \ket{\10,0}^0$ &
& 10 &
& $X_a^2 X_b^2 \ket{\10,0}^0$ &
& $\f_{(aabb)}$ &\cr
height2pt&\omit&&\omit&&\omit&&\omit&&\omit&\cr
\noalign{\hrule}
height3pt&\omit&&\omit&&\omit&&\omit&&\omit&\cr
& $\f_{(aabc)}$ &
& $X_a^2 X_b X_c \ket{\10,0}^0$ &
& 30 &
& $X_a^2 X_b X_c \ket{\10,0}^0$ &
& $\f_{(aabc)}$ &\cr
height2pt&\omit&&\omit&&\omit&&\omit&&\omit&\cr
\noalign{\hrule}
height3pt&\omit&&\omit&&\omit&&\omit&&\omit&\cr
& $\f_{(abcd)}$ &
& $X_a X_b X_c X_d \ket{\10,0}^0$ &
& 5 &
& $X_a X_b X_c X_d \ket{\10,0}^0$ &
& $\f_{(abcd)}$ &\cr
height2pt&\omit&&\omit&&\omit&&\omit&&\omit&\cr
\noalign{\hrule}
height3pt&\omit&&\omit&&\omit&&\omit&&\omit&\cr
& $\vf_{a(\b\g)}$ &
& $X_a Y_\b Y_\g \ket{\10,0}^0_{a=3,4,5}$ &
& 3 &
& $X_\a \ket{\frc34,\frc34}^4_{a=3,4,5}$ &
& $\e^{\b\g}\vq_a$ &\cr
height2pt&\omit&&\omit&&\omit&&\omit&&\omit&\cr
\noalign{\hrule}
height3pt&\omit&&\omit&&\omit&&\omit&&\omit&\cr
& $\e^{\a\b} \vq_{(aab)}$ &
& $X_a^2 X_b \ket{\inv4,\inv4}^4$ &
& 20 &
& $X_a^2 X_b Y_\a Y_\b \ket{\10,0}^0$ &
& $\e^{\a\b}\e^{AB} \vq_{(aab)}$ &\cr
height2pt&\omit&&\omit&&\omit&&\omit&&\omit&\cr
\noalign{\hrule}
height3pt&\omit&&\omit&&\omit&&\omit&&\omit&\cr
& $\e^{\a\b} \vq_{(abc)}$ &
& $X_a X_b X_c \ket{\inv4,\inv4}^4$ &
& 10 &
& $X_a X_b X_c Y_\a Y_\b \ket{\10,0}^0$ &
& $\e^{\a\b}\e^{AB} \vq_{(abc)}$ &\cr
height3pt&\omit&&\omit&&\omit&&\omit&&\omit&\cr}
\hrule height1pt}$$
\vskip5pt
\noindent
{\bf Table 3}: The Koszul and the Landau-Ginzburg orbifold
               rendition of the $(c,c)$-states for the models
               $\cal Q$ and $\Tw{\cal Q}$. Recall,
               $a,b..=1,\ldots,5$ and $\a,\b,A,B=1,2$; different
               labels mean different values ($a\ne b$ {\it etc}.).}
\ifx\answ\bigansw\vfill
            \else\vfill\goodbreak\vglue0mm\vfill\fi
\vbox{
$$\vbox{\offinterlineskip
\hrule height1pt
\halign{
&\vrule width1pt#&\strut~~#\hfil~~&\vrule#
                  &\strut~~#\hfil~~&\vrule#
                   &\strut~~#\hfil~~&\vrule#
                    &\strut~~#\hfil~~&\vrule#
                     &\strut~~#\hfil~~&\vrule width1pt#\cr
height2pt&\omit&&\omit&&\omit&&\omit&&\omit&\cr
&\hfil$(a,a)$&&\hfil$(c,c)$&&\hfil Ramond &
&\hfil$(a,c)$&&\hfil$(c,a)$&\cr
height2pt&\omit&&\omit&&\omit&&\omit&&\omit&\cr
\noalign{\hrule\vskip1pt\hrule}
height3pt&\omit&&\omit&&\omit&&\omit&&\omit&\cr
&$\ket{\1{-}3,{-}3}^0$&
&$\2{\2{\ket{\10,0}^0}}$&
&$\ket{\1{-}\frc32,{-}\frc32}^0$&
&$\ket{\1{-}3,0}^1$&
&$\ket{\10,{-}3}^7$&\cr
height3pt&\omit&&\omit&&\omit&&\omit&&\omit&\cr
\noalign{\hrule}
height3pt&\omit&&\omit&&\omit&&\omit&&\omit&\cr
&$\ket{\1{-}3,0}^1$&
&$\ket{\10,{+}3}^1$&
&$\ket{\1{-}\frc32,{+}\frc32}^1$&
&$\ket{\1{-}3,{+}3}^2$&
&$\ket{\10,0}^0$&\cr
height3pt&\omit&&\omit&&\omit&&\omit&&\omit&\cr
\noalign{\hrule}
height3pt&\omit&&\omit&&\omit&&\omit&&\omit&\cr
&$\ket{\1{-}1,{-}2}^2$&
&$\ket{\1{+}2,{+}1}^2$&
&$\ket{\1{+}\frc12,{-}\frc12}^2$&
&$\2{\2{\ket{\1{-}1,{+}1}^3}}$&
&$\ket{\1{+}2,{-}2}^1$&\cr
height3pt&\omit&&\omit&&\omit&&\omit&&\omit&\cr
\noalign{\hrule}
height3pt&\omit&&\omit&&\omit&&\omit&&\omit&\cr
&$\ket{\1{-}\frc53,{-}\frc83}^3$&
&$\ket{\1{+}\frc43,{+}\frc13}^3$&
&$\ket{\1{-}\frc16,{-}\frc76}^3$&
&$\2{\2{\ket{\1{-}\frc53,{+}\frc13}^4}}$&
&$\ket{\1{+}\frc43,{-}\frc83}^2$&\cr
height3pt&\omit&&\omit&&\omit&&\omit&&\omit&\cr
\noalign{\hrule}
height3pt&\omit&&\omit&&\omit&&\omit&&\omit&\cr
&$\ket{\1{-}1,{-}2}^4$&
&$\ket{\1{+}2,{+}1}^4$&
&$\ket{\1{+}\frc12,{-}\frc12}^4$&
&$\2{\2{\ket{\1{-}1,{+}1}^5}}$&
&$\ket{\1{+}2,{-}2}^3$&\cr
height3pt&\omit&&\omit&&\omit&&\omit&&\omit&\cr
\noalign{\hrule}
height3pt&\omit&&\omit&&\omit&&\omit&&\omit&\cr
&$\ket{\1{-}2,{-}1}^5$&
&$\ket{\1{+}1,{+}2}^5$&
&$\ket{\1{-}\frc12,{+}\frc12}^5$&
&$\ket{\1{-}2,{+}2}^6$&
&$\ket{\1{+}1,{-}1}^4$&\cr
height3pt&\omit&&\omit&&\omit&&\omit&&\omit&\cr
\noalign{\hrule}
height3pt&\omit&&\omit&&\omit&&\omit&&\omit&\cr
&$\ket{\1{-}\frc83,{-}\frc53}^6$&
&$\ket{\1{+}\frc13,{+}\frc43}^6$&
&$\ket{\1{-}\frc76,{-}\frc16}^6$&
&$\ket{\1{-}\frc83,{+}\frc43}^7$&
&$\ket{\1{+}\frc13,{-}\frc53}^5$&\cr
height3pt&\omit&&\omit&&\omit&&\omit&&\omit&\cr
\noalign{\hrule}
height3pt&\omit&&\omit&&\omit&&\omit&&\omit&\cr
&$\ket{\1{-}2,{-}1}^7$&
&$\ket{\1{+}1,{+}2}^7$&
&$\ket{\1{-}\frc12,{+}\frc12}^7$&
&$\ket{\1{-}2,{+}2}^8$&
&$\ket{\1{+}1,{-}1}^6$&\cr
height3pt&\omit&&\omit&&\omit&&\omit&&\omit&\cr
\noalign{\hrule}
height3pt&\omit&&\omit&&\omit&&\omit&&\omit&\cr
&$\ket{\10,{-}3}^8$&
&$\ket{\1{+}3,{+}0}^8$&
&$\ket{\1{+}\frc32,{-}\frc32}^8$&
&$\ket{\10,0}^0$&
&$\ket{\1{+}3,{-}3}^7$&\cr
height3pt&\omit&&\omit&&\omit&&\omit&&\omit&\cr}
\hrule height1pt}$$
\vskip5pt
\noindent
{\bf Table 4}: The Ramond and Neveu-Schwarz vacua of the
               Landau-Ginzburg orbifold $\cal M$.}

\vfill
\eject

 %
\listrefs

 %
\bye